\documentclass[manuscript]{aastex}

\usepackage{amsmath}
\usepackage{natbib}
\usepackage{footnote}
\usepackage{color}

\begin{document}

%% LaTeX will automatically break titles if they run longer than
%% one line. However, you may use \\ to force a line break if
%% you desire.

\title{EXONEST: Bayesian Model Selection \\ Applied to the Detection and Characterization \\
of Exoplanets Via Photometric Variations}

%% Use \author, \affil, and the \and command to format
%% author and affiliation information.
%% Note that \email has replaced the old \authoremail command
%% from AASTeX v4.0. You can use \email to mark an email address
%% anywhere in the paper, not just in the front matter.
%% As in the title, use \\ to force line breaks.

\author{Ben Placek and Kevin H. Knuth \altaffilmark{1}}
\affil{Physics Department, University at Albany (SUNY),
    Albany, NY 12222}
\email{bplacek@albany.edu}
\email{kknuth@albany.edu}

\author{Daniel Angerhausen}
\affil{Department of Physics, Applied Physics, and Astronomy, Rensselear Polytechnic Institute, Troy, NY 12180 }
\email{daniel.angerhausen@gmail.com}

%% Notice that each of these authors has alternate affiliations, which
%% are identified by the \altaffilmark after each name.  Specify alternate
%% affiliation information with \altaffiltext, with one command per each
%% affiliation.

\altaffiltext{1}{Department of Informatics, University at Albany (SUNY), Albany, NY 12222}
%\altaffiltext{2}{Society of Fellows, Harvard University.}
%\altaffiltext{3}{present address: Center for Astrophysics,
%    60 Garden Street, Cambridge, MA 02138}
%\altaffiltext{4}{Visiting Programmer, Space Telescope Science Institute}
%\altaffiltext{5}{Patron, Alonso's Bar and Grill}

%% Mark off your abstract in the ``abstract'' environment. In the manuscript
%% style, abstract will output a Received/Accepted line after the
%% title and affiliation information. No date will appear since the author
%% does not have this information. The dates will be filled in by the
%% editorial office after submission.

\begin{abstract}
EXONEST is an algorithm dedicated to detecting and characterizing the photometric signatures of exoplanets, which include reflection and thermal emission, Doppler boosting, and ellipsoidal variations. Using Bayesian inference, we can test between competing models that describe the data as well as estimate model parameters. We demonstrate this approach by testing circular versus eccentric planetary orbital models, as well as testing for the presence or absence of four photometric effects. In addition to using Bayesian model selection, a unique aspect of EXONEST is the potential capability to distinguish between reflective and thermal contributions to the light curve. A case-study is presented using Kepler data recorded from the transiting planet KOI-13b.  By considering only the non-transiting portions of the light curve, we demonstrate that it is possible to estimate the photometrically-relevant model parameters of KOI-13b.  Furthermore, Bayesian model testing confirms that the orbit of KOI-13b has a detectable eccentricity.

\end{abstract}

%% Keywords should appear after the \end{abstract} command. The uncommented
%% example has been keyed in ApJ style. See the instructions to authors
%% for the journal to which you are submitting your paper to determine
%% what keyword punctuation is appropriate.


\section{Introduction}

 Exoplanets are known to cause a variety of photometric effects, which collectively can be used for both exoplanet detection and characterization. The Kepler Space Telescope (Kepler) has reached a level of photometric precision and temporal coverage that allows for the detection of such effects \citep{Jenkins&Doyle:2003}. A portion of the observed flux variation originates directly from the exoplanet itself as both a reflected light component and thermal emission.  Additional effects originate from the influence of the exoplanet on its host star.  These include Doppler boosting, or beaming, caused by the radial velocity variations due to the stellar wobble, as well as variations in flux caused by the ellipsoidal distortion of the star, which is induced by the planetary tidal forces. These effects make it possible to detect \emph{non-transiting} planets, which are expected to account for a large subset of the extant exoplanet population. Using Bayesian inference one can estimate the values of the physical parameters on which these effects depend. In addition, by computing the Bayesian evidence a variety of models can be tested, some of which may either allow for, or neglect, these effects.

%The Kepler Space Telescope (Kepler) has reached a level of photometric precision and temporal coverage that allows for the detection of a variety of photometric effects caused by planets closely orbiting their host stars \citep{Jenkins&Doyle:2003}.
 % This includes light both reflected and emitted from the atmosphere or surface of a planet, as well as photometric effects originating from a planet's affect on the host star, such as relativistic Doppler beaming (or boosting) due to an induced stellar wobble, and ellipsoidal tidal distortions of the star itself.

%This fact makes it possible to detect \emph{non-transiting} planets, which are expected to account for a large subset of the extant exoplanet population.

The BEER algorithm, published by Faigler and Mazeh \citep{Faigler&Mazeh:2011}, is a novel way of detecting such planets via BEaming, Ellipsoidal variations, and Reflected light effects.  This has proved to be an efficient method for detecting companions, which include a good number of non-eclipsing binaries as well as Jupiter mass companions \citep{Faigler+etal:2012,Shporer+etal:2011}.  The BEER model assumes that planets have circular orbits and treats each effect as a sinusoid.  In short, this allows the algorithm to estimate the orbital period of the companion, and the amplitudes of the three effects (Beaming, Ellipsoidal, and Reflection).  %The main drawback is that it leaves out the possibility of eccentric orbits which have a wide range of
%photometric signatures that can deviate drastically from the sinusoidal form.
While it is likely that most detections will be short period Jupiter-mass planets in circular orbits, planets closely orbiting their host stars with significant eccentricities have been found  and are interesting from a planetary formation standpoint \citep{Matsumura+etal:2008}.

Additional advantages of considering each of these photometric effects include potentially breaking the $M_p \sin i$ degeneracy associated with other detection techniques, as well as providing information about scattering and emissive properties of exoplanet atmospheres \citep{Seager+etal:2000,Charbonneau+etal:2002,Hood+etal:2008,Rowe+etal:2008}. In our algorithm, we include separate models for both reflected light and thermal emissions with the aim of potentially differentiating between these two effects in a photometric data set.  This typically cannot be done for hot-Jupiters, since the majority have circularized orbits, which result in thermal phase curves that are identical in shape to those of reflected light. However, for eccentric orbits, these two effects are potentially separable due to the fact that there is a $1/r^2$ drop off in the reflected light that does not appear in the thermal emissions.

Our methodology is based on \emph{Bayesian Model Selection}, which is accomplished by computing the Bayesian evidence (hereafter called the evidence) for each of a set of models.  The evidence allows one to compare the probability of one model to another.  Similar methods have been employed to test between different models of multi-planet systems using radial velocity data \citep{Feroz+etal:2011,Gregory:2011} as well as for finding secondary eclipses in CoRoT light curves \citep{Parviainen+etal:2013}.  In this application to Kepler data, by turning photometric effects on-and-off, one can determine the probability with which each effect, or combination of effects, contributes to the total recorded photometric signal.  In situations where an exoplanet model is found to require several of these effects, or particular combinations of these effects, this tends to increase confidence that the data originates from an actual exoplanet.  In a very real sense, each photometric effect can be conceived to represent an independent detection technique.

\section{Bayesian Model Selection}

To make inferences from data, we rely on Bayes' Theorem
\begin{equation}
P(\theta_M|D,M) = \frac{P(\theta_M|M) P(D|\theta_M,M)}{P(D|M)}
\end{equation}
where $M$ represents the hypothesized model, $D$ represents the recorded data, and $\theta_M$ represents the set of parameters belonging to model $M$.  The prior probability, $P(\theta_M|M)$, quantifies what is known about the parameters $\theta_M$ before considering the data $D$. The likelihood function, $P(D|\theta_M,M)$, quantifies the probability that the specific model and its hypothesized parameter values could have produced the data.  The probability in the denominator, $P(D|M)$, is the evidence, which represents the probability that the model (irrespective of its parameter values) could have given rise to the data. Finally, the posterior probability, $P(\theta_M|D,M)$, quantifies what is known about the model and its parameter values after the data have been considered.  In this sense, Bayes' Theorem acts as an update rule that takes what is known before the data are considered (prior probability) and modifies it with the ratio of two data-dependent terms (likelihood and evidence) resulting in the posterior probability.  The posterior probability is critical for estimating the model parameter values $\theta_M$; whereas the evidence allows one to assess the probability that the model $M$ describes the data.

The evidence is calculated by integrating, or marginalizing, over all model parameter values:
\begin{equation}
P(D|M) = Z = \int P(\theta_M | M) P(D|\theta_M,M) \, d\theta_M.
\end{equation}
For this reason, the evidence, which is often called the marginal likelihood, can be considered to be a prior-weighted average of the likelihood.
Since models with more parameters have a lower prior probability (unit probability is spread over a larger space), with the likelihoods being equal, the Bayesian evidence naturally favors the simpler model.  In this sense Bayes' theorem is often envisioned to implement a form of Occam's Razor.

Imagine that we have two models, $M_1$ and $M_2$, that we would like to compare and test by applying them to a data set.  To determine which model is favored by the data, we compute the posterior odds ratio by dividing the posterior probability of one model by the posterior probability of the other:
\begin{equation}
\frac{P(M_1|D,I)}{P(M_2|D,I)} = \frac{P(M_1|I)}{P(M_2|I)}\frac{P(D|M_1)}{P(D|M_2)} = K \frac{P(M_1|I)}{P(M_2|I)}.
\end{equation}
%KK-
where $I$ represents one's prior information about the problem and $K$ is the Bayes' factor, which represents the ratio of the model evidences.
%-KK
The ratio of the model priors is often set to one implying no prior preference to either model.
As a result, the evidence quantifies the probability of a model given the data, so that the Bayes' factor enables one to compare the probabilities of a pair of models.
%As a result, the Bayes' factor is proportional to the ratio of the probabilities of each model given the data.

To compute the evidences, we must integrate the product of the prior and the likelihood over the entire, usually multi-dimensional, parameter space.  In practice, such computations can be performed numerically by means of the Nested Sampling algorithm \citep{Sivia&Skilling:2006}, or a version of nested sampling called MultiNest \citep{Feroz&Hobson:2008,Feroz+etal:2009,Feroz+etal:2013}, which is well-equipped to handle multimodal probability distributions.  Another benefit of the Nested Sampling algorithm, and its cousin MultiNest, is that they provide posterior samples, which allows one to compute parameter estimates and uncertainties in those estimates in addition to model-testing.
%% In this section, we use  the \subsection command to set off
%% a subsection.  \footnote is used to insert a footnote to the text.

%% Observe the use of the LaTeX \label
%% command after the \subsection to give a symbolic KEY to the
%% subsection for cross-referencing in a \ref command.
%% You can use LaTeX's \ref and \label commands to keep track of
%% cross-references to sections, equations, tables, and figures.
%% That way, if you change the order of any elements, LaTeX will
%% automatically renumber them.

%% This section also includes several of the displayed math environments
%% mentioned in the Author Guide.

\section{Modeling Photometric Variability}

Accurate modeling of exoplanet-induced photometric variability requires considering at least three different mechanisms \citep{Loeb&Gaudi:2003,Faigler&Mazeh:2011}, each of which depends on the orbit of the planet. Therefore, the first step is to generate an orbit from the set of hypothesized model parameters.  We find that the most efficient way of doing this is to iterate the mean, eccentric, and true anomalies following % KK added
%-KK
the method used in %% BP - added another reference for this
Brown \citep{Brown:2009} and Mislis et al. \citep{Mislis+etal:2011}.
This can be performed by iterating the following equations
over the elapsed time from $t=0$ to an arbitrary $t = t_{end}$:
%%BP - NEW VERSION OF EQUATIONS
\begin{equation}
M(t) = M_0 + \frac{2 \pi t}{T}
\end{equation}
\begin{equation} \label{eq:eccentric_anomaly}
E(t) = M(t) + e \sin E(t)
\end{equation}
\begin{equation} \label{eq:true-anomaly}
\tan \frac{\nu(t)}{2} = \sqrt{\frac{1+e}{1-e}} \tan \frac{E(t)}{2}
\end{equation}
where $M_0$ is the initial mean anomaly,
$T$ is the orbital period, $e$ is the orbital eccentricity, and $M(t)$, $E(t)$ and $\nu(t)$ are the mean, eccentric, and true anomalies, respectively. The eccentric anomaly (\ref{eq:eccentric_anomaly}) is given by the transcendental equation which is solved via the Newton-Raphson method with a stopping criterion given by $|E_i - E_{i+1}| < 10^{-8}$. The distance $r$ between the star and planet is calculated using the
%three
eccentric and true %KK added
anomalies:
\begin{equation} \label{eq:distance}
r(t) = a ( 1 - e \cos E(t)) = \frac{a (1 - e^2)}{1 + e \cos \nu(t)}.
\end{equation}

Given $r(t)$, one can calculate the position of the planet in Cartesian coordinates $(x, y, z)$ at any time given the orbital inclination $i$, the argument of periastron $\omega$, the true anomaly $\nu(t)$, and the line of nodes $\Omega$ %(eqn (8)).

\begin{equation}
\begin{pmatrix} x(t) \\ y(t) \\ z(t) \end{pmatrix}
= r(t) \begin{pmatrix}
\cos\Omega \cos(\omega+\nu(t)) - \sin\Omega \sin(\omega + \nu(t)) \cos i
\\ \sin\Omega \cos(\omega+\nu(t)) + \cos\Omega \sin(\omega+\nu(t)) \cos i
\\ \sin(\omega+\nu(t)) \sin i
 \end{pmatrix}.
\end{equation}
Since the line of nodes serves only to rotate the orbit in the reference plane
about the line-of-sight, %KK added
it does not affect the observed flux variation.
We can therefore simplify the calculations by setting $\Omega$ equal to zero \citep{Brown:2009}, which results in

\begin{equation}
\begin{pmatrix} x(t) \\ y(t) \\ z(t) \end{pmatrix}  = r(t) \begin{pmatrix}
\cos(\omega+\nu(t))
\\ \sin(\omega+\nu(t)) \cos i
\\ \sin(\omega+\nu(t)) \sin i
 \end{pmatrix}.
\end{equation}

The vector $\vec{r}(t) = x(t)\,\hat{x} + y(t)\,\hat{y} + z(t)\,\hat{z}$ gives the position of the planet, so that the unit vector $\hat{r}(t) = \frac{\vec{r}(t)}{r(t)}$ points from the star to the planet.  Defining the
%KK line of site
line-of-site to be $\hat{r'} = \hat{z}$, %KK added
one can calculate the phase angle $\theta(t)$ of the planet

\begin{align} \label{eq:phase_angle}
\theta(t) &= \arccos( \hat{r}(t) \cdot \hat{r'} ) \nonumber \\
%&= \arccos\Big( \frac{\vec{r}(t) \cdot \hat{z}}{r(t)} \Big) \nonumber \\
&= \arccos\Big( \frac{z(t)}{r(t)} \Big) \nonumber \\
&= \arccos( \sin(\omega+\nu(t)) \sin i ).
\end{align}
Accurate prediction of the position of the planet as a function of time is critical to obtaining accurate predictions of the photometric effects.

\subsection{Reflected Light}

Given the position of the planet at any given time, one can model the time-series of the expected photometric effects.  To model reflected light, we begin by assuming that the star radiates isotropically and that the planet acts as a Lambertian sphere. The total light reflected from the planet is found by integrating over the illuminated surface visible along the line-of-sight.

The
%amount of light reflected off of %  %KK reworded
infinitesimal luminosity reflected from % KK reworded
a surface element, $\hat{n}\,dA$, on a planet is given by
\begin{equation}
dL_p = A_{eff} F_0 \, \hat{n} \cdot \hat{r} \, dA
\end{equation}
where $A_{eff}$ is an effective albedo, $\hat{n} = \sin \alpha \cos\beta \, \hat{x} + \sin \alpha \sin \beta \, \hat{y} + \cos \beta \, \hat{z}$ is the unit normal vector to the planetary surface, $\hat{r} = \sin \theta(t) \, \hat{x} + \cos \theta(t) \, \hat{y}$ is the
unit %KK insertion
orbital radius vector, the surface area element on the planet is $dA = {R_p}^2 \sin \alpha \, d\alpha \, d\beta$ and $F_0$ is the incident stellar flux given by
\begin{equation} \label{eq:stellar-flux}
F_0 = \frac{L_\star}{4 \pi r(t)^2},
\end{equation}
where $L_\star$ is the stellar luminosity and $r(t)$ is the potentially time-varying orbital separation between star and planet. %KK reworded
Taking the dot-product
%and integrating we get  %KK deleted
\begin{equation}
\hat{n} \cdot \hat{r} = \sin \alpha \sin(\theta(t) + \beta)
\end{equation}
and integrating %KK insertion
\begin{align}
L_p &= \int \,dL_p \nonumber \\
&= A_{eff} F_0 \int_0^\pi \sin^2\alpha \, d\alpha \int_{-\theta}^{\pi-2\theta} \sin(\theta(t) +\beta) \, d\beta \label{eq:luminosity-integral}
\end{align}
we get %KK insterion
%Integrating on $0 < \alpha < \pi$ and $-\theta<\beta < \pi-2\theta $ we obtain
\begin{equation}
L_p(t) = \frac{A_{eff} F_0 \pi {R_p}^2}{2} \left( 1 + \cos \theta(t) \right).
\end{equation}
This expression has units of luminosity, which is made more explicit by substituting (\ref{eq:stellar-flux}) so that %KK reworded
\begin{equation} \label{eq:planetary-luminosity}
L_p(t) = \frac{A_{eff}}{8} \frac{{R_p}^2}{r(t)^2} L_\star \left( 1 + \cos \theta(t)\right).
\end{equation}

The flux of the planet observed from Earth is
\begin{equation}
F_p(t) = \frac{L_p(t)}{4 \pi d^2},
\end{equation}
which given the luminosity in (\ref{eq:planetary-luminosity}), can be rewritten as
\begin{equation}
F_p(t) = \frac{\frac{A_{eff}}{8} \frac{{R_p}^2}{r(t)^2} L_\star}{4 \pi d^2} \left( 1 + \cos \theta(t)\right)
\end{equation}
where $d$ is the distance from Earth to the planet.
It is common practice to normalize with respect to the stellar flux received at Earth
\begin{equation}
F_\star = \frac{L_\star}{4 \pi d^2}
\end{equation}
so that
\begin{align}
\frac{F_p(t)}{F_\star} &= \frac{4 \pi d^2}{L_\star}  \frac{\frac{A_{eff}}{8} \frac{{R_p}^2}{r^2} L_\star}{4 \pi d^2} \left( 1 + \cos \theta(t)\right) \nonumber \\
&= \frac{A_{eff}}{8} \frac{{R_p}^2}{r(t)^2} \left(1 + \cos \theta(t) \right).
\end{align}

From definitions found in \citep{seager:2010}, the albedo $A_{eff}$ can be written as
\begin{equation}
A_{eff} = g A_s
\end{equation}
where $A_s$ is the spherical albedo, and $g$ is a correction factor to account for
%kkedit
%the planet not scattering equally into $4\pi {R_p}^2$
anisotropic scattering from the planet.
This correction factor can be written in terms of the spherical and geometric albedos \citep{seager:2010}
\begin{equation}
g = \frac{4A_g}{A_s},
\end{equation}
which gives the final result for the normalized reflected component of the planetary flux
\begin{equation} \label{eq:photometric-reflection}
\frac{F_p(t)}{F_\star} = \frac{A_g}{2} \frac{{R_p}^2}{r(t)^2} \left(1 + \cos \theta(t) \right).
\end{equation}

\subsection{Thermal Emission}

Since the majority of planets detectable by photometric variations will be close to their host stars $(a < 0.1 \mbox{AU})$, it is expected that thermal emission will play a role in the total flux emanating from the planet. For example, thermal emissions have been detected in light curves of certain transiting planets such as the hot Jupiter, HAT-P-7b \citep{Borucki+etal:2009}. It is thought to be nearly impossible to distinguish the thermal photons from reflected photons \citep{Cowan&Agol:2011} in the case of close-in hot Jupiters. This should be the case for circular orbits where both reflected light and thermal emissions vary sinusoidally. However, for sufficiently eccentric orbits the reflected light curve can deviate significantly from a sinusoid providing the opportunity to distinguish the two photometric effects. The ability to separate thermal emissions from reflected light will depend on the eccentricity of the orbit as well as the signal to noise of the data.
%kkedit
%We use an approach similar to the derivation of reflected light.

The infinitesimal thermal luminosity from a surface element on the dayside of a planet, denoted $d L_{Th,d}$, is given by
\begin{equation}
d L_{Th,d} =  F_p(T_d) \hat{n} \cdot \hat{r} \, dA
\end{equation}
where $F_p(T_d)$ is the thermal flux from the dayside of the planet. Integrating over the surface of the planet, as in (\ref{eq:luminosity-integral}), yields the thermal luminosity of the dayside
\begin{equation}
L_{Th,d}(t) = \frac{F_p(T_d) \pi {R_p}^2}{2} \left(1+\cos \theta(t) \right).
\end{equation}
The flux received at Earth is given by
\begin{align}
F_{Th,d}(t) &= \frac{L_{Th,d}(t)}{4 \pi d^2} \nonumber \\
&= \frac{F_p(T_d) {R_p}^2}{8d^2} \left( 1 + \cos\theta(t) \right).
\end{align}
Normalizing by the stellar flux gives
\begin{align}
\frac{F_{Th,d}(t)}{F_\star} &= \frac{ \frac{ F_p(T_d) {R_p}^2}{8d^2} }{ \frac{L_\star}{4 \pi d^2} } \left(1+\cos \theta(t) \right) \nonumber \\
&= \frac{F_p(T_d) \pi {R_p}^2}{2 L_\star} \left( 1 + \cos \theta(t) \right).
\end{align}
Substituting the stellar luminosity $L_\star = F_\star(\pi R_\star^2)$ results in
\begin{equation}
\frac{F_{Th,d}(t)}{F_\star} = \frac{1}{2} \left( \frac{R_p}{R_\star} \right)^2  \left( 1 + \cos \theta(t) \right) \frac{F_p(T_d)}{F_\star}.
\end{equation}
%%KK-
The stellar flux detected by the Kepler bandpass \citep{VanCleve&Caldwell:2009} is found by integrating, over all possible wavelengths $\lambda$, the product of the spectral radiance of a blackbody
\begin{equation}
B(T) = \frac{2hc^2}{\lambda^5}\frac{1}{e^\frac{hc}{\lambda k_B T} -1}
\end{equation}
evaluated at the effective temperature of the star $T= T_{eff}$
and the Kepler response function $K(\lambda)$ \citep{VanCleve&Caldwell:2009}
\begin{equation}
F_\star = \int B(T_{eff})K(\lambda) \, d\lambda.
\end{equation}
The planetary thermal flux detected by the Kepler bandpass is found similarly using the dayside and nightside temperatures $T_d$ and $T_n$ so that
%%These equations yield the following form for the dayside and nightside contributions to the thermal phase curve
%%-KK
\begin{equation} \label{eq:photometric-thermal-day}
\frac{F_{Th,d}(t)}{F_\star} = \frac{1}{2}(1 + \cos\theta(t)) \left( \frac{R_p}{R_\star} \right)^2 \frac{ \int B(T_d)K(\lambda) \, d\lambda}{\int B(T_{eff}) K(\lambda) \, d\lambda}
\end{equation}
and
\begin{equation} \label{eq:photometric-thermal-night}
\frac{F_{Th,n}(t)}{F_\star} =\frac{1}{2} (1 + \cos(\theta(t) - \pi)) \left( \frac{R_p}{R_\star} \right)^2 \frac{ \int B(T_n)K(\lambda) \, d\lambda}{\int B(T_{eff}) K(\lambda) \, d\lambda},
\end{equation}
where $R_\star$ is the radius of the star.  These integrals can be performed by numerical integration.

\subsection{Doppler Boosting}

The Doppler boosting component of the photometric variation originates from a relativistic effect that occurs because of the stellar wobble induced by the planet.  As the star moves toward
%KK towards is more common in Britain.  Both toward and towards are correct
an observer there is an increase in the observed flux, and as it recedes the observed flux decreases.  There is also a boosting component from the reflected light from the planet, however the amplitude of the reflected light is so small compared to the total stellar flux that the stellar boosting far outweighs the planetary boosting despite the fact that the planet is traveling much faster around the center of mass.  This effect can be quantified by \citep{Rybicki&Lightman:1979}
\begin{equation}
F_{boost}(t) = F_{\star} \left( \frac{1}{\gamma (1 - \beta \cos \theta(t))} \right)^4
\end{equation}
where $\gamma^{-1} = \sqrt{1-\beta^2}$, $\beta = \frac{v}{c}$ where $c$ is the speed of light, and $F_{\star}$ is the stellar flux in the reference frame of the star. Since even in the most extreme cases the stellar radial velocities will be on the order of $10^2$ to $10^3 m s^{-1}$ ($\beta \sim 10^-5$), we can use the non-relativistic limit \cite{Loeb&Gaudi:2003}.  Acknowledging the time dependence of the effect, we can approximate the boosting component of the flux by
\begin{equation} \label{eq:photometric-boosting}
\frac{F_{boost}(t)}{F_\star} =  1 + 4\beta_r(t)
\end{equation}
 %%BP -
where $\beta_r$ is the component of the stellar velocity along the line-of-sight $\hat{z}$, more commonly referred to as the radial velocity.  This is given by
\begin{equation}
\beta_r(t) = \frac{v_z(t)}{c}
\end{equation}
where
\begin{equation}
v_z(t) = K(\cos(\nu(t) + \omega) + e \cos \omega)
\end{equation}
such that $\nu(t)$ is the true anomaly given by (\ref{eq:true-anomaly}) and $K$ is the radial velocity semi-amplitude, which
in units of $m s^{-1}$ can be found by
\begin{equation}
K = 28.435 \left( \frac{T}{1yr} \right)^{-\frac{1}{3}} \frac{M_p \sin i}{M_J} \left( \frac{M_\star}{M_\odot} \right)^{-\frac{2}{3}}.
\end{equation}
where $T$ is the orbital period in units of years, $M_p$ is the mass of the planet in Jupiter masses, and $M_\star$ is the mass of the star in solar masses.
It is important to note that both the orbital inclination $i$ and the mass of the planetary companion $M_p$ play unique roles in the boosting effect.  This implies that in the case where the photometric variations have a significant boosting component, there is the potential to estimate both the inclination and the mass of the planetary companion, which is impossible from reflected light alone.

\subsection{Ellipsoidal Variations}
Ellipsoidal variations arise from stellar distortion due to planetary gravitational tidal forces.  To first order, the star is shaped like a prolate spheroid with the semi-major axis pointing approximately toward the planet, with a potential lag that could also be modeled.  This results in a periodic fluctuation of the observed stellar flux that is equal to half the period of the planetary orbit as the visible cross section of the star changes. Since it would be computationally expensive to model this effect precisely,
we instead model these variations using the estimated amplitude given by Loeb and Gaudi  \citep{Loeb&Gaudi:2003},
with a trigonometric modification \citep{Kane&Gelino:2012}
\begin{equation} \label{eq:photometric-ellipsoidal}
\frac{F_{ellip}(t)}{F_\star} = \beta \frac{M_p}{M_\star} \left( \frac{R_\star}{r(t)} \right)^3 [\cos^2(\omega+\nu(t)) + \sin^2(\omega+\nu(t))\cos^2i]
\end{equation}
where $M_p$ and $M_\star$ are the masses of the planet and star, respectively, $R_\star$ is the radius of the star, $r(t)$ is the distance from the star to the planet given by (\ref{eq:distance}), $\nu(t)$ is the true anomaly given by (\ref{eq:true-anomaly}), $\omega$ is the argument of periastron, and $\beta$ is the gravity darkening exponent given by
\begin{equation}
\beta = \frac{\log\Big(\frac{GM_{\star}}{R_{\star}^2}\Big)}{ \log T_{eff}  },
\end{equation}
where $T_{eff}$ is the effective temperature of the host star.
Since both the boosting and ellipsoidal variations both depend on the true anomaly $\nu(t)$ and the argument of periastron $\omega$, we can model these effects both in the case of circular and eccentric orbits.
%%-KK

\subsection{Combining Photometric Effects}

%%% Figure 1 A-C: Circular and Eccentric Orbits with Beaming and Ellipsoidal variations.
\begin{figure}[h!]
\centering
\includegraphics[ scale = .5 ]{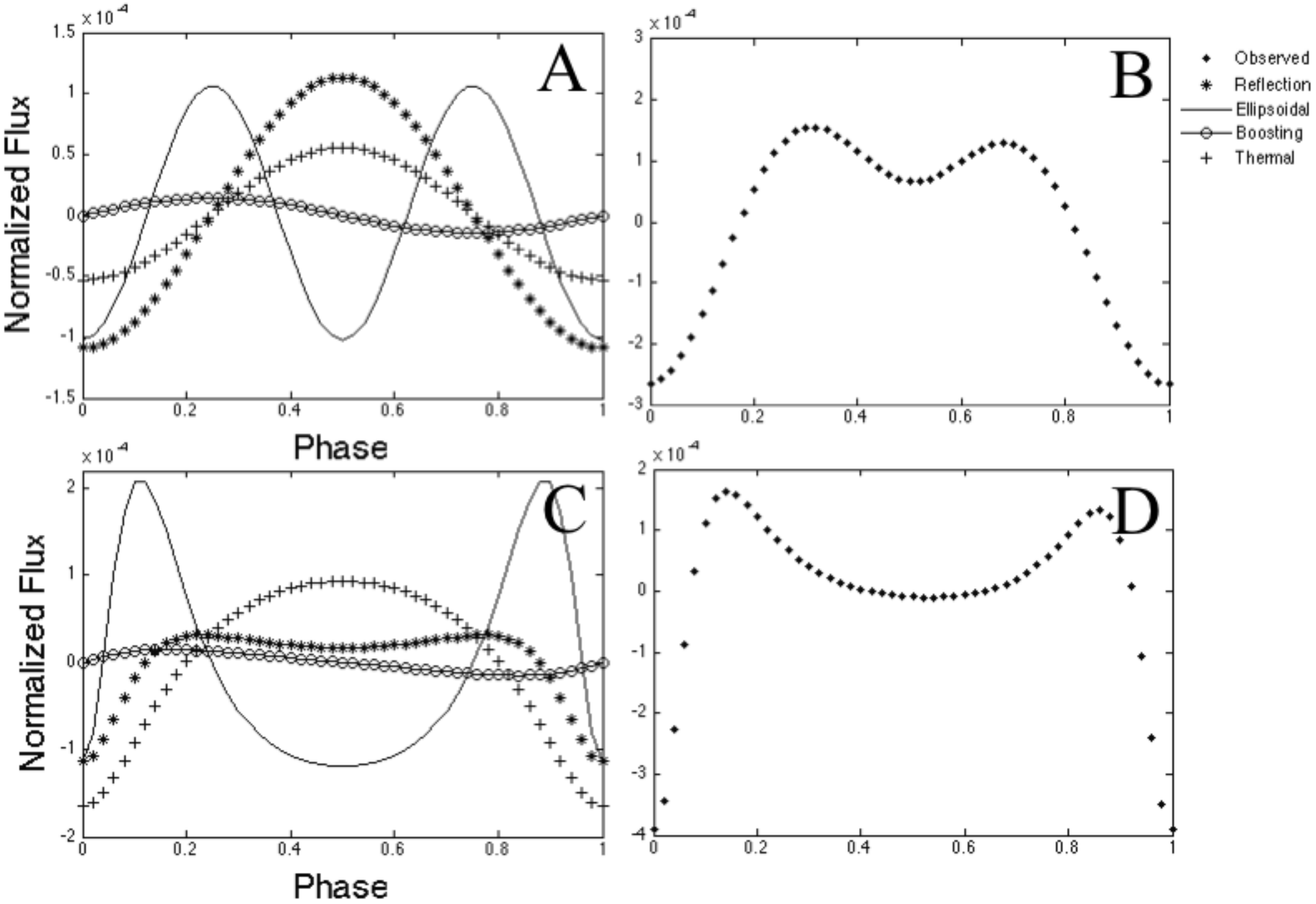}
\caption{(Left Column, A and C) An illustration of the reflected light, Doppler boosting, ellipsoidal variation, and thermal emission components.  (Right Column,B and D ) These effects combine to produce a net observed photometric variation. The top row (A and B) corresponds to a planet with a circular orbit while the bottom row (C and D) corresponds to planet with an eccentric (e = 0.3) orbit.  Thermal effects, which depend significantly on day- and nightside temperatures as well as the eccentricity of the orbit, cannot always be distinguished from reflected light.
 }
\end{figure}
The net photometric variation of a modeled exoplanetary system can be found by first computing the orbital position $\left(r(t), \theta(t) \right)$ of the planet, or planets, as a function of time as given by (\ref{eq:distance}) and (\ref{eq:phase_angle}).  This, combined with the model parameters describing the star and planet, can then be used to compute the components of photometric flux due to light reflected from the planet (\ref{eq:photometric-reflection}), Doppler boosting of the starlight (\ref{eq:photometric-boosting}), ellipsoidal variations in the shape of the star (\ref{eq:photometric-ellipsoidal}), and thermal emission contributions from the dayside (\ref{eq:photometric-thermal-day}) and nightside (\ref{eq:photometric-thermal-night}) of the planet.  By simply summing these photometric flux contributions, one can generate a predictive model of the observed photometric flux variations
\begin{equation}
F_{pred}(t) = F_\star \left(1 + \frac{F_p(t)}{F_\star} + \frac{F_{boost}(t)}{F_\star} + \frac{F_{ellip}(t)}{F_\star} + \frac{F_{Th,d}(t)}{F_\star} + \frac{F_{Th,n}(t)}{F_\star}\right).
\end{equation}
Often photometric timeseries are normalized by dividing the observed flux $F_{obs}$ by the mean flux $<F_{obs}>$ and then subtracting that mean so that
\begin{equation}
F_{norm}(t) = \frac{F_{obs}(t)}{<F_{obs}>} - <F_{obs}>.
\end{equation}
Since $<F_{obs}> \approx {F_\star}$, one can use the following predictive model with normalized photometric data
\begin{equation}
F_{normpred}(t) \approx \frac{F_p(t)}{F_\star} + \frac{F_{boost}(t)}{F_\star} + \frac{F_{ellip}(t)}{F_\star} + \frac{F_{Th,d}(t)}{F_\star} + \frac{F_{Th,n}(t)}{F_\star},
\end{equation}
with the understanding that this is only approximate since the interesting photometric effects do not necessarily average to zero and as a result also contribute to the $<F_{obs}>$.  Throughout the remainder of this paper, we will work with normalized data.

Figure 1 illustrates three photometric components, as well as the resulting observed signal, in the case of both a planet in a circular orbit (top) and the same planet in an eccentric orbit with $e = 0.3$ (bottom).
%%kkedit- I assume the parameters are the same for both the circular and eccentric case.  For this reason I moved epsilon out to avoid confusion.
The model parameters for this example are $\{M_\star = 2.05M_\odot, R_\star = 2.55 R_\odot, i = 80^o, \omega = 3\pi/2, M_0 = 0, R_p = 1.86 R_J, M_p = 8.5 M_J, T_{eff} = 8500, T_d = 3400K, T_n = 3000K\}$.  Consider a planet in a circular orbit (Figures 1A and B).  The reflected light varies sinusoidally at the same period as the orbit of the planet cycles through its phases with a maximum flux corresponding to the planet being in a full phase (depending, of course, on the orbital inclination).  On the other hand, the ellipsoidal variations oscillate sinusoidally at twice the orbital period with maxima occurring when the planet is in either a first quarter or last quarter phase.  This corresponds to the star being oriented so that its cross-sectional area along the line-of-site is at a maximum.  The boosting component varies with the period of the orbit, but is off by a phase from the reflected light since the maximum boosting occurs when the planet is in its first quarter phase and the star is moving toward the observer.

The photometric variations of an equivalent planet in an eccentric orbit ($e = 0.3$) are illustrated in Figures 1C and D.  The most dramatic difference is the fact that the reflected light time series is bimodal.  This occurs because the stellar flux received by the planet decreases with $1/r^2$ so that as the planet approaches apastron the reflected flux decreases.  The change of shape of the curves representing flux from ellipsoidal variations and boosting are directly due to the fact that it is an eccentric orbit rather than a circular orbit.

\section{Analysis, Methodology and Testing}{
\subsection{Model Testing}

Exoplanets can exhibit several different photometric effects.  However, each of these effects may be present to varying degrees, or possibly absent altogether, dependent on the exoplanet.  This implies that a careful analysis should consider a set of models, each consisting of a different subset of effects.  One of the unique features of EXONEST is its ability to both perform parameter estimation as well as statistical model testing.

Each of the models involves a different number of parameters depending on the photometric effects that are included.  Moreover, different model parameters have different ranges of possible values, so that different models possess parameter spaces of different dimensionalities and dimensions.  This affects the prior probability assigned to a model, and thus one's inferences.

Bayes' Theorem naturally weighs the ability of a model to describe the data against the complexity of that model as quantified by the volume of its parameter space.  Therefore, Bayes' favors the simpler of two models that predict the data equally well.  Bayesian model selection allows us to test whether specific photometric effects, such as reflection, boosting, ellipsoidal variations, or thermal emissions, play a major role in describing the data.  For example, if the boosting and tidal effects are negligible, a model that includes these effects will have a lower evidence and will be less favored.  On the other hand, if the reflected light variations are negligible, then the more complex models must describe the data very well in order to overcome the penalty imposed by having a larger parameter space.

When considering reflected light, the flux involves the product of the geometric albedo and the radius of the planet squared, $A_g {R_p}^2$, which together act as a single parameter affecting the amplitude of the cosine.  Thermal emissions depend only on the radius of the planet, and not on the geometric albedo.  In the case of a circular orbit, thermal emissions cannot be disentangled from reflected light.  For this reason, in this paper, we do not test for thermal emissions in the case of circular orbits, though we keep in mind that the photometric signal could still involve both reflected and thermally-emitted components.  The situation is expected to be more interesting in the case of highly-eccentric close-in orbits since the thermal emissions will behave differently from reflected light.  In this case it may be that the two effects can be disentangled.

\subsection{The Prior and Likelihood Function}
Bayesian inference requires us to assign both the prior probabilities of the model parameters and the likelihood of the data given the hypothesized parameter values.  In this application, we assign a uniform prior probability over a reasonable range to each of the model parameters as indicated in Table 1.  Such assignments can be changed as we add layers of sophistication to this methodology, and as more is learned about exoplanets in general.

One important point should be made involving the prior assigned to the orbital inclination.  Since one does not expect a relationship between the orientation of a planet's orbit and the orientation of the observer, the inclination angle is sampled from a uniform distribution on a sphere. To do this, one can either estimate the inclination using the arccosine of the uniform distribution from $[0,\frac{\pi}{2}]$ \citep{Brown:2009}, or one can estimate $\cos i$ using a uniform prior from $[0,1]$. In some cases, more informative priors may be employed for certain parameters. For instance, \citep{Kipping:2013} proposes the use of a beta-distribution for a prior on orbital eccentricity and \citep{Gregory&Fischer:2010} employ an eccentricity bias correction filter to account for situations with low signal-to-noise.

\begin{table}[h]
\label{tbl -1}
\begin{tabular}{l c c c c c}
  Parameter & Variable & & Interval & & Distribution \\
\hline
Orbital Period (Days) & T  &  & $[0.01,15]$ & & Uniform \\
Eccentricity & e & & $[0,1]$ & & Uniform \\
Stellar Mass $(M_\odot)$ & $M_s$  &  & Known & &  \\
Mean Anomaly $(rad)$ & $M_0$ & & $[0,2\pi]$ &  & Uniform \\
Arg. of Periastron $(rad)$& $\omega$ & & $[0,2\pi]$ & & Uniform \\
Orbital Inclination $(deg)$&  $i$ &  & $[0,\frac{\pi}{2}]$ & & Uniform on Sphere \\
Minimum Planetary Radius $(R_J)$ & $\sqrt{A_g}R_p$ & & $[10^{-4},4]$ & & Uniform \\
Planetary Radius $(R_J)$ & $R_p$ & & $[10^{-4},4]$ & & Uniform \\
Geometric Albedo	&	$A_g$ 	& &	$[0,1]$ &	&	Uniform \\
Stellar Radius $(R_\odot)$ & $R_\star$  & & Known & &  \\
Planetary Mass $(M_J)$ & $M_p $ & & $[.1,20]$ & & Uniform \\
Dayside Temperature (K) &$T_{d}$ & & $[0,6000]$ & & Uniform \\
Standard Deviation of Noise (ppm) & $\sigma$ & &$[10^{-6},10^{-4}]$ & & Uniform\\

\hline
 \end{tabular}
\caption{Prior Distributions for Model Parameters}

\end{table}

The likelihood function quantifies the degree to which one expects the photometric time-series predicted by the model to agree with the data.  Quantifying all of the information that one may possess about the system can be a difficult task since this would include accounting for the optics of the telescope, the dynamics of the CCD pixels, and all of the preprocessing that went into the available data.  Experience has shown that in many cases, it is often best to make simple assumptions---at least as a start.  In the case where the mean value of the signal and the expected squared deviation about the mean (variance) are the only two relevant parameters, the principle of maximum entropy dictates that the least biased assignment is a Gaussian likelihood.  Such an assignment accommodates the relevant parameters while ensuring that no other unintended effects are assumed.  However, in this situation, is not clear how best to handle the variance parameter.  In many cases, marginalizing over the unknown signal variance to obtain a Student-t likelihood \citep{Student:1908,Sivia&Skilling:2006} provides the most conservative estimate.  However, In our preliminary tests with synthetic data, we found that the Student-t likelihood adversely affected the model testing portion of the analysis.  For example, when testing a circular orbit model on synthetic data simulating a planet with an eccentric orbit, the long tails of the Student-t distribution were able to accommodate the extreme non-sinusoidal photometric variations associated with a planet in an eccentric orbit. As a result, we found that circular orbit models were consistently favored over the eccentric orbit models despite obvious photometric signatures of orbital eccentricity in the data.

With the signal variance, $\sigma^2$, either known or included as a model parameter, one can assign a Gaussian likelihood, which depends on the sum of the square differences between the Kepler photometric data value $d_i$ recorded at time $t_i$ and the photometric time-series at time $t_i$, $F_{M}(t_i) \equiv F_{M}(t_i,\theta_M)$, predicted by the model $M$ and its model parameters $\theta_M$
\begin{equation}
P(d_i | \theta_M, M) = \frac{1}{\sqrt{2 \pi \sigma^2}} \exp \left( {-\frac{\left(F_{M}(t_i) - d_i \right)^2}{2 \sigma^2}} \right).
\end{equation}
Assuming that the value of a given data point does not depend on the fact that previous data were recorded, implies that the joint likelihood for all of the recorded data can be written as a product of the likelihoods of the individual data values $D = \{d_1, d_2, \ldots, d_N\}$.  As a result the logarithm of the joint likelihood is a sum of the individual log likelihoods
\begin{align}
\log P(D | \theta_M, M) &= \log P(\{d_1, d_2, \ldots, d_N\} | \theta_M, M) \nonumber \\
&= -\frac{1}{2 \sigma^2} \sum_{i=1}^{N} \left( F_{M}(t_i) - d_i \right)^2 - \frac{N}{2} \log{2 \pi \sigma^2},
\end{align}
where using $i$ as an index enables one to easily accommodate gaps in the data due to removed transits or other observation discontinuities.  The last term on the right, $- \frac{N}{2} \log{2 \pi \sigma^2}$ , is not only constant with respect to the model parameters, but also independent of the model, and can be ignored making the computations more efficient.
%BP
However, if the signal variance is unknown and is incorporated as a model parameter, this term must be considered.
%BP

The product of the prior and the likelihood are proportional to the posterior probability of the model parameters given the data.  Here, since we have employed uniform priors, the posterior probability is proportional to the likelihood function.  To explore the posterior probability, we employ the MultiNest algorithm, which efficiently integrates the product of the prior and the likelilhood to obtain the evidence calculation while collecting samples, which can be used to estimate the model parameters.  The result is that one can simultaneously perform Bayesian model testing and Bayesian parameter estimation.

\subsection{Analysis of Synthetic Data}
In order to better understand the capabilities of EXONEST, we tested the algorithm in a series of experiments using synthetic datasets derived from the photometric model described in Section 3.  The first set of experiments involved testing the ability of EXONEST to distinguish circular and eccentric orbits via model selection as well as estimating the variance of the noise as an additional model parameter.  The circular and eccentric synthetic data describes 10 days worth of measurements of a close-in hot Jupiter akin to KOI-13b in orbits with $e=0.0$ and $e=0.2$, respectively.  In both cases the geometric albedo of the hypothetical planet was taken to be $A_g = 0.15$.  Gaussian-distributed noise with a variance of $\sigma = 23.1 \times 10^{-6}$ was added to each time-series.  Both models were tested on each dataset including reflection, boosting, and ellipsoidal variations.  Thermal flux variations were neglected in these simulations.
With regard to model testing, a model is said to be favored if it has a higher (or less negative) evidence value than another model.  This assumes no prior preference for either model.  Evidence values are typically expressed as the natural logarithm of the evidence, denoted $\ln Z$, to accommodate the extremely large range of values typically encountered.  As a result, a difference in log evidence values indicates a factor of probability proportional to a power of the natural number $e \approx 2.718$.
Log evidences for each dataset are shown in Table 2.  In both cases the correct model had the larger log evidence value. %insert table number here and elaborate on the results.

\begin{table}[h]
\centering
\label{tbl-2}
\begin{tabular}{l c c}

	 & Eccentric Data & Circular Data \\

\hline
			
Eccentric Model & $\bf{4046.70}$	& $3988.40$ \\
Circular Model &	$3956.70$	& $\bf{3990.80}$ \\
 Null 		&	$3849.00$	&$3749.20$ \\
\hline

\end{tabular}

\caption{MultiNest log evidences ($\ln Z$) for synthetic datasets.  The model most favored to describe the data is in bold. In the case of the synthetic circular orbit (Circular Data), the correct Circular Model was approximately $exp(2.4)$ times more probable than the incorrect Eccentric Model. Even more significant results were obtained in the case of the synthetic eccentric orbit (Eccentric Data) where the correct Eccentric Model was approximately $exp(90)$ times more probable than the incorrect Circular Model.}

\end{table}

\begin{figure}
\centering
\includegraphics[scale=.45]{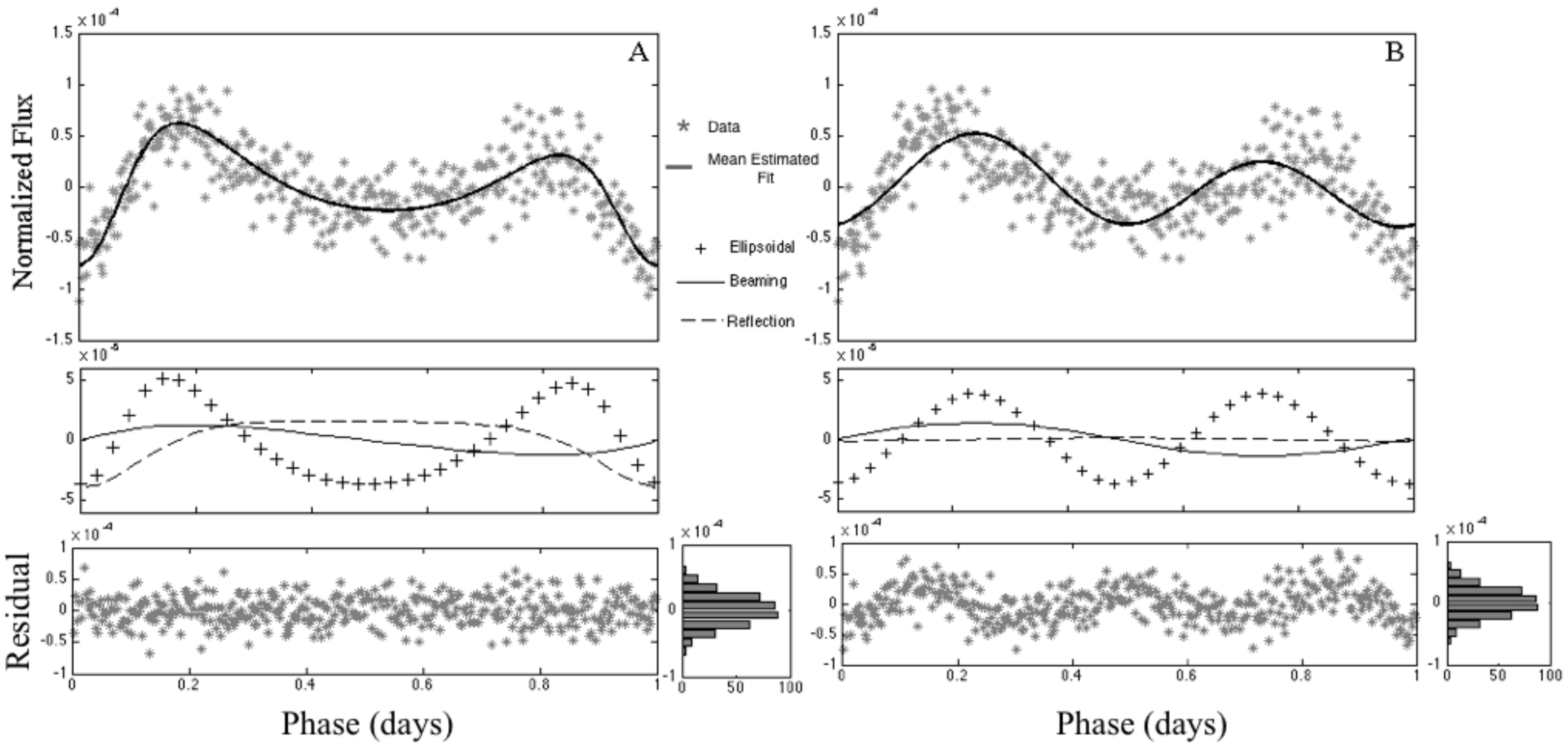}
\caption{Phase folded synthetic eccentric data with fits for eccentric (A) and circular (B) models including reflected light, Doppler beaming, and ellipsoidal variations.  The circular model clearly leaves un-modeled structure in the residuals. }
\end{figure}

Parameter estimates for both the circular and eccentric models applied to the eccentric dataset are shown in Table 3 and fits to the eccentric dataset are shown in Figure 2.  EXONEST was able to estimate the correct variance of the noise in the eccentric case showing the ability to handle datasets for Kepler candidates in which the noise level is unknown.  EXONEST significantly over estimated the noise variance in the case of the circular model indicating a poor fit, which is to be expected since the synthetic data actually describes an eccentric orbit.  This implies that the algorithm is treating the parts of the data that it cannot fit as noise.  It is also worth noting that in the case of the circular model, the estimated planetary radius was extremely small. This is due to the fact that the dominant component of the flux is the ellipsoidal variation to the stellar shape (as seen in Figures 2A and B); whereas, reflected light is the smallest effect.
The parameter denoted as $Rp_{true}$ is an estimate of the true radius of the planet taking the planetary albedo into account.  To obtain this value, we assumed that the planetary albedo was known to be $0.15 \pm 0.1$.  This relatively conservative prior, with a relatively large uncertainty, is the reason for the relatively large uncertainty in the true planetary radius.  In the next section, we describe the estimation of the planetary radius in more detail and show how to apply the procedure to cases where the true radius of the planet has been estimated from transit events.

\begin{table}[h]
\centering
\label{tbl-3}
\begin{tabular}{l c c c}
& \multicolumn{1}{c}{Eccentric } & \multicolumn{1}{c}{Circular } \\
\hline
Parameter		& 	Mean				& 	Mean			& 	Actual \\
T 			&$1.7630 \pm 0.0012$		& $1.7630 \pm 0.0016$ &	$1.7637$		    \\
$i$ 			&$80.80 \pm 6.45$			& $77.29 \pm 9.01$ 	&	$87.10$		\\
$M_0$ 		&$0.03 \pm 0.02$			& $4.79 \pm 0.04$		&	$0.00$		\\
$\sqrt{A_g}R_p$ 		&$0.73 \pm 0.05$			& $0.16 \pm 0.06$		& $0.77$\\
$M_p$ 		&$9.42 \pm 0.52$			&$10.50 \pm 0.80$		&	$8.35$		\\
e 			&$0.19 \pm 0.01$			& \nodata 			&	$0.20$		\\
$\omega$ 		&$4.74 \pm 0.04$			& \nodata				&	$4.71$		\\
$\sigma$ 		&$23.20 \pm 0.80$ppm 		&$30.20 \pm 0.77$ppm		&	$23.10$	ppm	\\
$Rp_{true}$	&$2.05 \pm 0.54$			&$8.50e-05 \pm 2.20e-05$ &  $2.00$		\\
\hline
\end{tabular}

\caption{MultiNest parameter estimates for the synthetic datasets.}

\end{table}

The second set of experiments was designed to determine how well EXONEST can estimate dayside temperatures when the thermal flux variations are included in the model.   There should be a threshold defined by the Kepler Response Function below which thermal flux variations will not be detected.  In addition, if the dayside and nightside temperatures are similar, the amplitude of the thermal flux variations will be small and difficult to detect.  For this experiment, 10 datasets describing the eccentric synthetic planet from the first two tests were created each with a different dayside temperature ranging from $T_d=3200$K to $T_d =5000$K  in $200$K increments, and with a nightside temperature fixed at $3000$K.

By examining Figure 3, it can be seen that nonzero nightside flux variations act to decrease the amplitude of the signal, and shift the mean upwards.  This effect could equivalently be explained by increasing the orbital inclination, which would decrease the amplitude of the oscillation, and by slightly increasing the mean flux of the star, which would increase the mean of the observed overall planetary flux.  This implies that there is an inherent degeneracy in the model involving these parameters, which can only be resolved by obtaining additional relevant information.  At this point in time, given the nature of the Kepler data, we resolve this degeneracy by setting the nightside temperature to zero in our model, which implicitly assumes that most planetary nightsides do not significantly radiate in the Kepler bandpass.  We should note that using two or more separate spectral channels would resolve the degeneracy.
The results from these simulations are shown in Figure 4.

\begin{figure}[h]
\centering
\includegraphics[scale=.5]{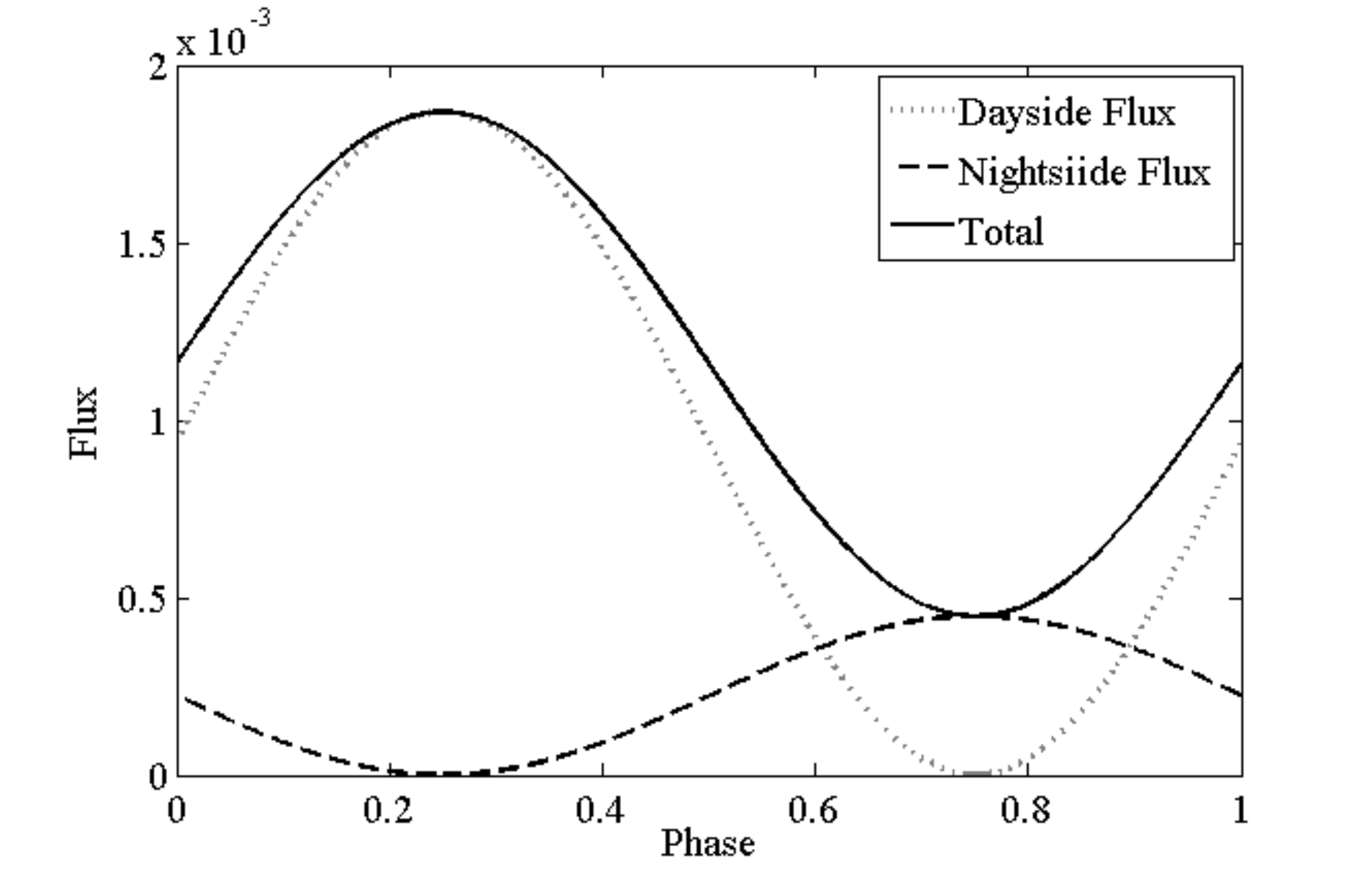}
\caption{Depiction of thermal flux variations. The gray dotted line represents the dayside flux, the black dashed line is the nightside flux, and the solid black line is the total observed flux. Thermal flux variations from the nightside act to shift the mean total signal upwards, while decreasing the total signal amplitude. This will also occur if the inclination and stellar flux are increased (thus decreasing the amplitude and increasing the mean).}
\end{figure}

\begin{figure}[h]
\centering
\includegraphics[ width = 170mm]{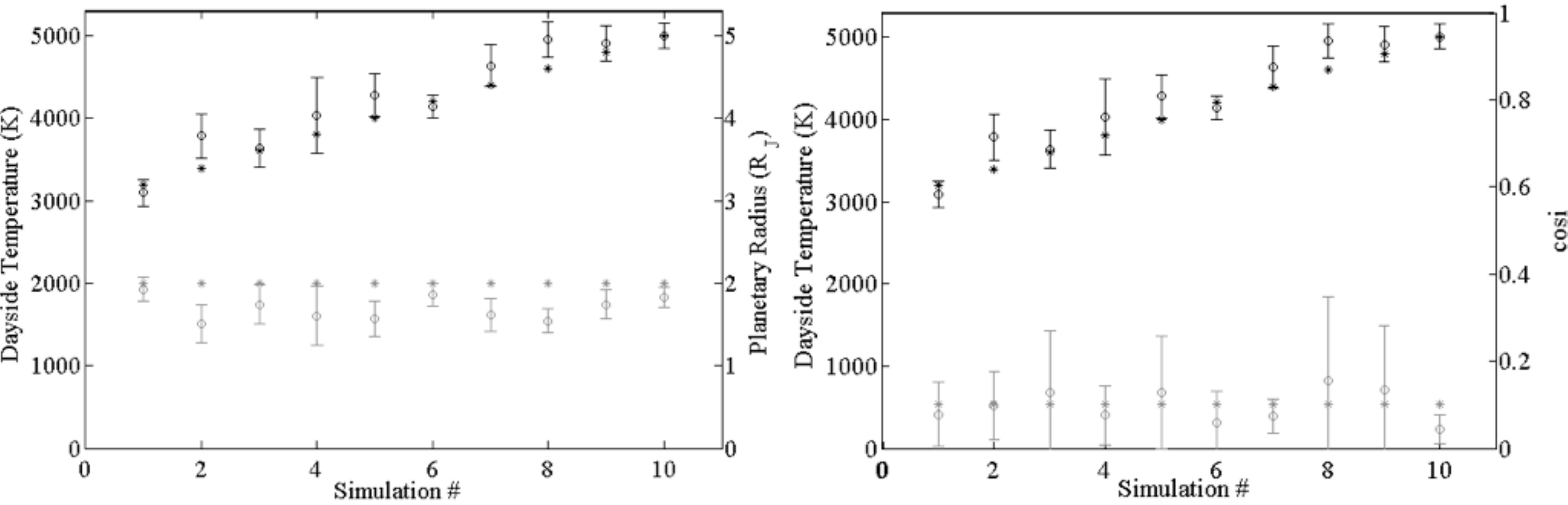}
\caption{MultiNest estimates of the dayside temperatures, $T_d$, (black) along with planetary radius (gray;left) and $\cos i$ (gray;right) from synthetic datasets involving eccentric orbits ($e=0.2$). The black and gray asterisks represent true parameter values.}
\end{figure}

The third set of experiments was designed to determine how eccentricity affects the ability of EXONEST to differentiate thermal emission from reflected light.  To demonstrate this, 10 datasets each with different eccentricites ranging from $e=0.00$ to $e=0.45$ in $0.05$ increments while keeping all other model parameters identical to the previous examples.  MultiNest was run 30 times on each dataset for a total of 300 runs so that the uncertainties in the log evidence could be estimated.  Figure 5 shows the mean log evidences for two different models applied to each of the 10 datasets.  The first model shown in black includes thermal flux variations, while the other, shown in gray, neglects them.  We found that for planets with eccentricities less than $e \sim 0.3$, EXONEST cannot definitively distinguish (at a $2\sigma$ level) between reflected light and thermal emission.  It should be noted that this limit governs ones ability to distinguish between the two models --- not the ability to estimate model parameter values as was illustrated in Figure 4, which involved a planet in an eccentric orbit with $e=0.2$.  Since in many cases, in the Kepler dataset, these two fluxes cannot be distinguished without further information, it may be more appropriate to refer to them as a single effect --- planetary flux.

Finally, in order to better understand degeneracies in the model, we have explored the log-likelihood probability landscape. Figure 7 shows two-dimensional slices through the log-likelihood probability landscape at the position of the correct solution (black crosses).  In this case synthetic data was used with parameters similar to those of KOI-13b (Note that these parameters are identical to those used in Experiments 1-3).  Of particular interest are the plots involving orbital inclination ($\cos i$), dayside temperature ($T_d$), planetary radius ($R_p$), and planetary mass ($M_p$) since they are all relatively flat.  This would imply that large uncertainties will be associated with these parameters in simulations.  As can be seen from (\ref{eq:photometric-thermal-day}), the planetary radius and dayside temperature both act to change the amplitude of the thermal flux. This degeneracy results in a relatively flat region in the $R_p$ vs. $T_d$ landscape.  The argument of periastron ($\omega$) and the initial mean anomaly ($M_0$) determine the shape and phase of the wave-form and thus can be estimated very precisely.  This is manifested as sharp peaks in the probability landscape for all of the plots involving these two angles.  It should be noted that these two-dimensional slices through parameter space were taken at the correct solution, so the geometric albedo was taken to be $A_g = 0.15$.  This was not assumed in the simulations performed on the data for experiment 1 where MultiNest estimated the joint parameter $\sqrt{A_g}R_p$.  A similar examination of the probability landscape for the Kepler observations of KOI-13b will be presented in the next section.

\begin{figure}[h]
\centering
\includegraphics[scale=.5]{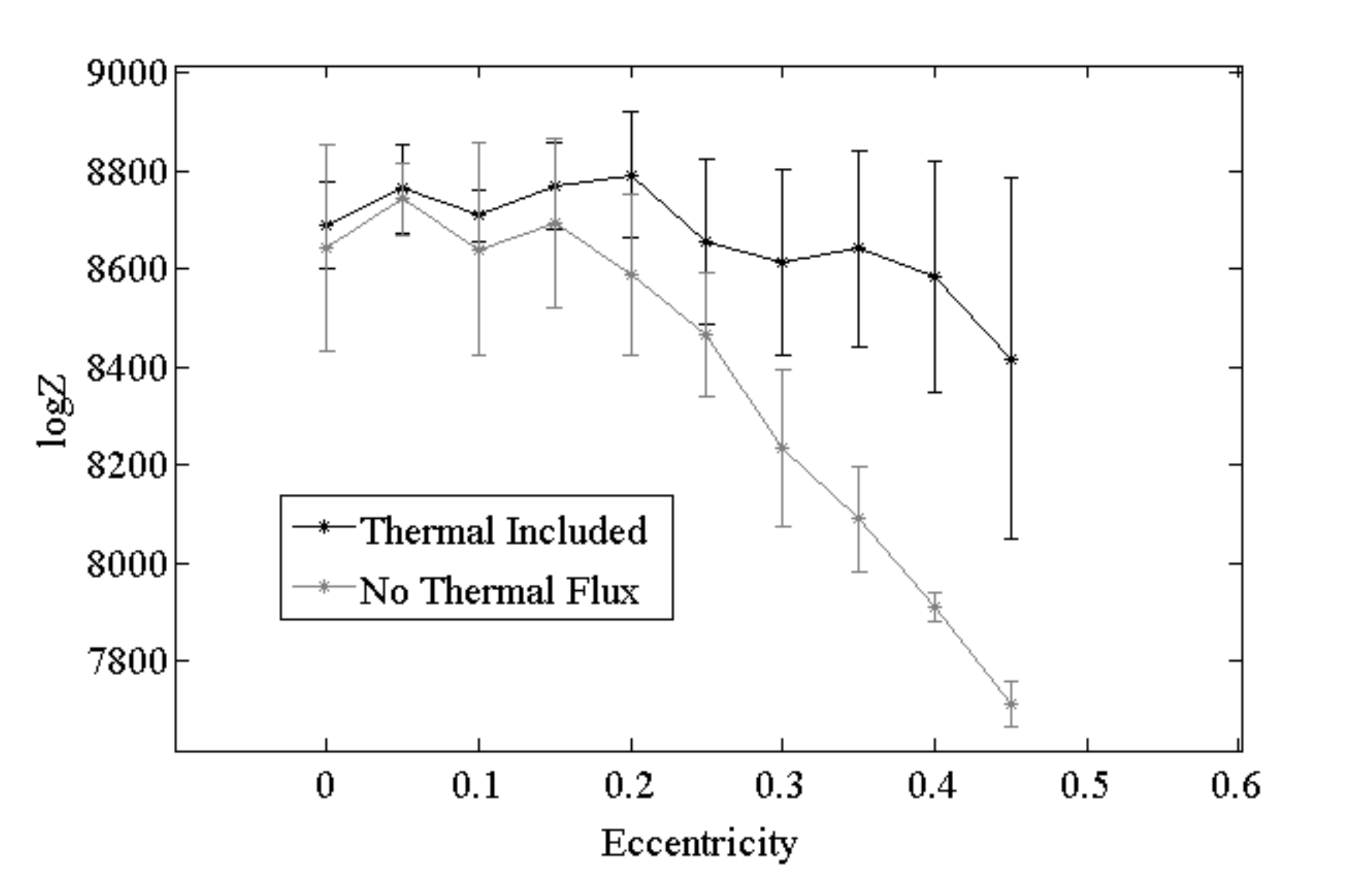}
\caption{Log evidence values obtained by applying a model that neglects thermal flux (black) and a model that includes thermal flux (gray) to the synthetic datasets of varying eccentricity.  It can be seen that one cannot definitively distinguish the two models (and therefore thermal and reflected flux) for eccentricities $< 0.3$.}
\end{figure}

\section{Analysis of KOI-13b}

KOI-13b is one of the largest transiting planets found to date. Using the BEER algorithm, Shporer et al. \citep{Shporer+etal:2011} demonstrated that KOI-13b is detectable from its photometric signal alone. To demonstrate our methodology, we apply 18 different photometric models to 121 days of out-of-transit data (4187 data points) from the first, second, and third quarters of the Kepler mission.  Each of these models represents different combinations of phometric effects, such as reflected light, Doppler boosting, ellipsoidal variations, and thermal emissions applied to either a circular orbit or an eccentric orbit.  In addition, we also tested the null model, which assumes that there is no planet around the star and that the star radiates with a constant flux.  In the case of KOI-13b, parameter estimation and log evidence calculations were all under fifteen minutes in duration. Previous studies of KOI-13b suggest that the planet is in a close-in circular orbit ($e = 0 \pm 0.05$, \citep{Mislis&Hodgkin:2012}), and as such, it induces detectable ellipsoidal variations as well as Doppler boosting \citep{Mislis&Hodgkin:2012,Shporer+etal:2011,Faigler&Mazeh:2011}.

We take the stellar radius, $R_{\star}$ and the effective temperature $T_{eff}$ to be known parameters.  These can be obtained from the Kepler Input Catalogue (KIC) estimates \citep{Latham+etal:2005}.
However, here we have taken them to be the estimates made by Szab{\'o} et al. \citep{Szabo+etal:2011} using both photometric and spectroscopic observations ($R_{\star} = 2.55 R_\odot, T_{eff} = 8500K$).  In order to significantly speed up computation time, we also phase-fold the data on the accepted period of $1.7637$ days.
 The MultiNest algorithm was then used to obtain both model evidences and parameter estimates, which are presented in Tables 4 and 5, respectively.  Each MultiNest simulation was performed using 100 live samples along with a stopping criterion of $tol = 0.1$.  The algorithm is terminated once the estimated evidence contribution from the current set of live samples is below this tolerance value \citep{Feroz+etal:2013}.  It should be noted that increasing the number of samples used by MulitNest increases the quality of the log evidence estimate.  However, this results in a significant increase in the computation time.  For this reason, we use 100 samples and run MultiNest $30$ times to ensure stability in the log evidence estimate and its uncertainties.

%Obtained log evidence values (Table 4) clearly indicate that the three most probable models are an eccentric orbit with  (a) reflection, boosting, and ellipsoidal variations ($\ln Z = 37847 \pm 20$), (b) thermal, boosting, and ellipsoidal variations ($\ln Z = 37865 \pm 30$), and (c) reflection, thermal, boosting, and ellipsoidal variations ($\ln Z = 37851 \pm 45$).  We were able to determine uncertainties for the log evidence values by performing the analysis 30 times for each of these three models.  It is because of these uncertainties that we cannot definitively distinguish between these three models.  What this does mean is that Doppler boosting and ellipsoidal variations are important effects that describe the data, and that reflected light and thermal emission cannot be distinguished in this case as predicted by our simulations on synthetic datasets.
 Obtained log evidence values (Table 4) clearly indicate that the two most probable models are an eccentric orbit with thermal emission, boosting, and ellipsoidal variations ($\ln Z = 37\,764.0 \pm 8.3$) and reflection, thermal emission, boosting, and ellipsoidal variations ($\ln Z = 37\,765.0 \pm 0.9$).  Based on the uncertainties of the log evidences for these two models we can not distinguish between them.  Thus we conclude that thermal emission is the more significant effect in that it dominates the planetary flux.  The second and third most probable models were the eccentric orbit with reflection, boosting, and ellipsoidal variations ($\ln Z = 37\,748.0 \pm 1.1$), and both circular orbits with reflection, boosting, and ellipsoidal variations ($\ln Z = 37\,703.0 \pm 1.1$) and thermal, boosting, and ellipsoidal variations ($\ln Z = 37\,703.0 \pm 0.5$).  This latter result was to be expected since in the circular case, the reflection and thermal effects have the same signature.

Finally, the fourth and fifth most probable models were circular and eccentric orbits with reflection and ellipsoidal variations.  It is not surprising that the circular model with only reflection and ellipsoidal variations has a high log evidence since the phase-folded light curve (Figure 6) clearly shows variations at the same frequency as the orbit (reflection/thermal), and the double-peaked waveform with half of the orbital frequency (ellipsoidal variations).
  As far as testing between circular and eccentric orbits,  a more standard evaluation technique was performed, involving the goodness-of-fit, which is found by the sum of the squared residuals for the two models.  Again, the eccentric model was found to be favored since the sum of the squared residuals (RSS) for the eccentric model is $RSS = 3.45e-06$; whereas for the circular model the sum of the squared residuals is $RSS = 3.8e-06$, which is almost 10\% larger.  The two models with the largest log evidence also had the lowest $RSS$ values.  The eccentric model including thermal emission, Doppler boosting, and ellipsoidal variations had $RSS = 3.3674e-06$; whereas the eccentric model that included those three effects and reflected light gave an $RSS = 3.3686e-06$.

\begin{table}[h]
\centering
\label{tbl-4}
\begin{tabular}{l c c}
			Model		&  Circular ($\ln Z$)	&  Eccentric ($\ln Z$)  \\
\hline
Refl. Only				& $37\,108.0 \pm 0.4 (5)$	& $37\;659.0 \pm 5.4 (7)$ \\

Boost Only			& $36\,970.0 \pm 4.0 (5)$     & $37\,166.0 \pm 1.9 (7)$\\

Ellips. Only			& $36\,555.0 \pm 0.5 (5) $    & $37\,581.0 \pm 0.4 (7)$\\			

Refl. + Boost. 			& $37\,108.0 \pm 0.5 (6)$ 	& $37\,670.0 \pm 2.9 (8)$ \\

Refl. + Ellips.			& $37\,701.0 \pm 0.5 (6)$	& $37\,704.0 \pm 2.7(8)$ \\

Boost. + Ellips. 		& $36\,577 \pm 0.8  (5)$	& $37\,634.0 \pm 2.8 (7)$  \\

Refl. + Boost. + Ellips.	& $37\,703 \pm 1.1 (6)$	       & $37\,748 \pm 1.1 (8)$ \\

Therm. + Boost + Ellips. & $37\,703 \pm 1.1 (8)$ & $\bf{37\,764 \pm 8.3} (10)$ \\

Refl. + Boost. + Ellips. + Therm. & \nodata & $\bf{37\,765.0 \pm 0.9} (10)$ \\

\hline
 Null & \multicolumn{2}{c}{$36\,143 \pm 1.0 (1)$} \\
\end{tabular}

\caption{MultiNest log evidences ($\ln Z$) for 18 different models applied to the photometric signal of KOI-13b.  The models most favored to describe the data are in bold and the number of model parameters for each model are given in parentheses.}

\end{table}

%% NEW VERSION OF PARAMETER ESTIMATES TABLE
\begin{table}[h!]  \label{Table:Estimates}

\centering
\label{tbl-5}
\begin{tabular}{c c c c}
%& \multicolumn{1}{c}{Eccentric } & \multicolumn{1}{c}{Circular }  \\
& Eccentric & Circular &  \\
 \hline
 Parameter  & Mean   & Mean  &  \bf{Accepted}  \\
\hline
  e       			& $0.034 \pm 0.003$  		& \nodata   			& $\bf{< 0.05}^b$ \\
  $M_s$    		& \nodata   				& \nodata				& $\bf{2.05}^b$  \\
 $\omega$   		&  $3.57 \pm 0.13$  		& \nodata				& \nodata  \\
 $M_0$	 		& $1.02\pm 0.13$    		& $5.05 \pm 0.16$ 		& \nodata  \\
$\sqrt{A_g}R_p$ 	&$0.748 \pm 0.015$		& $0.693 \pm 0.005$  	& \nodata \\
$A_g$			& 	\nodata				&\nodata				& \nodata \\
$R_p$ 			& $1.98 \pm 0.31$			&$1.95 \pm 0.51$		&$\bf{1.86 \pm 0.003}^c$ \\
i 				& $81.37 \pm 5.23 $ 		& $85.30 \pm 1.72$  	& $\bf{85.135^{+.097}_{-0.063}}^d$ \\
$R_\star$ 		&  \nodata   				& \nodata				& $\bf{2.55}^b$\\
$M_p$ 			& $8.35 \pm 0.43$  		& $6.35 \pm 0.11$  	& $\bf{8.30 \pm 1.25}^e$\\
$T_{d}$ 			& \nodata					& \nodata				& \nodata \\
$\sigma$ (ppm)	&  $28.00 \pm 0.50 $ 		& 	$33.00\pm 0.50$ 	&\nodata\\
\hline
$\ln Z$ & $37\,748 \pm 1.1 $ &  $37\,703  \pm 1.1 $  &  \\

\end{tabular}
\tablenotetext{a}{\citep{Shporer+etal:2011}}
\tablenotetext{b}{\citep{Szabo+etal:2011}}
\tablenotetext{c}{\citep{Borucki+etal:2011}}
\tablenotetext{d}{\citep{Esteves+etal:2013} }
\tablenotetext{e}{\citep{Mislis&Hodgkin:2012}}
\caption{MultiNest parameter estimates for KOI-13b. The Circular and Eccentric models included Reflected Light, Doppler Boosting, and Ellipsoidal Variations.  The bottom row lists the log evidence, $\ln Z$, for each of the three models.  The uncertainties in log evidence were obtained by running MultiNest 30 times on each dataset.  Estimates of $R_p$ were calculated post-simulation by the process explained in section 4.4.}
\end{table}

%% NEW VERSION OF PARAMETER ESTIMATES TABLE
\begin{table}[h!]  \label{Table:ThermalEstimates}

\centering
\label{tbl-6}
\begin{tabular}{c c c c}
%& \multicolumn{1}{c}{Eccentric } & \multicolumn{1}{c}{Circular }  \\
& w/Reflection & No Reflection &  \\
 \hline
 Parameter  & Mean   & Mean  &  \bf{Accepted}  \\
\hline
  e       			& $0.061 \pm 0.006$  		& $0.062 \pm 0.005$   			& $\bf{< 0.05}^b$ \\
  $M_s$    		& \nodata   				& \nodata				& $\bf{2.05}^b$  \\
 $\omega$   		&  $3.43 \pm 0.11$  		& $3.42 \pm 0.10$		& \nodata  \\
 $M_0$	 		& $1.22 \pm 0.11$    		& $1.23 \pm 0.10$ 		& \nodata  \\
$\sqrt{A_g}R_p$ 	&\nodata					& \nodata			  	& \nodata \\
$A_g$			& $0.09 \pm 0.10$		&\nodata				& \nodata \\
$R_p$ 			& $0.60 \pm 0.11$			&$2.00 \pm 0.60$		&$\bf{1.86 \pm 0.003}^c$ \\
i 				& $81.37 \pm 5.23 $ 		& $81.5 \pm 5.7$  		& $\bf{85.135^{+.097}_{-0.063}}^d$ \\
$R_\star$ 		&  \nodata   				& \nodata				& $\bf{2.55}^b$\\
$M_p$ 			& $7.10 \pm 0.32$  		& $7.10 \pm 0.60$  	& $\bf{8.30 \pm 1.25}^e$\\
$T_{d}$ 			& $5263.0 \pm 394$			& $3492.4 \pm 340.4$	& $\bf{3\,724^{+5}_{-6}}^d$ \\
$\sigma$ (ppm)	&  $29.00 \pm 0.50 $ 		& $29.00\pm 0.50$ 	&\nodata\\
\hline
$\ln Z$ & $37\,764 \pm 8.3 $ &  $37\,765  \pm 0.9 $  &  \\

\end{tabular}
\tablenotetext{a}{\citep{Shporer+etal:2011}}
\tablenotetext{b}{\citep{Szabo+etal:2011}}
\tablenotetext{c}{\citep{Borucki+etal:2011}}
\tablenotetext{d}{\citep{Esteves+etal:2013} }
\tablenotetext{e}{\citep{Mislis&Hodgkin:2012}}
\caption{MultiNest parameter estimates for KOI-13b. Both models assumed eccentric orbits while including the following photometric effects: thermal emission, Doppler boosting, and ellipsoidal variations (2nd column), and reflected light, thermal emission, Doppler boosting, and ellipsoidal variations.  The bottom row lists the log evidence, $\ln Z$, for both models.  The uncertainties in log evidence were obtained by running MultiNest 30 times on each dataset.  }
\end{table}

% As can be seen from both Table 5 and Figure 6, the thermal component seems to have been suppressed.  The estimated dayside temperature of KOI-13b was $T_d = 1387.00 \pm 794.00 K$.  This implies that the eccentricity of KOI-13b is too near circular to enable one to disentangle the thermal flux from the reflected flux based on photometric variations alone. The thermal model also could not precisely estimate the orbital inclination giving a mean and standard deviation of $i = 69.47 \pm 2.43$. In this case, we know this to be wrong because KOI-13b is transiting and would be well beyond transiting at that inclination.  The geometric albedo $A_g$ was estimated poorly as well.  An alternative to estimating $A_g$ is to model it given the planet-star separation distance

Figure 6 compares the folded data points to the predictions made by the circular and eccentric models for reflection, boosting, and ellipsoidal variations (A and B).  The model that included reflected light, Doppler boosting, and ellipsoidal variations predicts KOI-13b to have a mass of $M_p = 8.35 \pm 0.43M_J$, a minimum radius of $\sqrt{A_g}R_p = 0.748 \pm 0.015 R_J$, and a slightly eccentric orbit with $e = 0.034 \pm 0.003$. The beta distribution was implemented as an alternative prior on eccentricity in order to determine if the preference for eccentric orbits was due to the uniform prior. As proposed by \citep{Kipping:2013}, the beta distribution well-describes the distribution of eccentricities for $\sim 400$ known exoplanets. For our analysis, we used $A = 0.867$ and $B = 3.03$ for the shape parameters. This alternative prior did not significantly change our parameter estimates or the uncertainties, yielding an eccentricity of $e = 0.034 \pm 0.002$, and $\ln Z = 37850 \pm 20$.
Given the planetary radius of KOI-13b, $R_p = 1.860 \pm 0.003R_J$ estimated by \citep{Borucki+etal:2011} using transit data, we can derive an estimate of the geometric albedo.  Based on the provided information, we assigned a Gaussian probability density function for $R_p$ with a mean given by $\overline{R}_p = 1.860$ and a standard deviation given by $\sigma_{R_p} = 0.003$
\begin{equation}
Pr(R_p | I) = \frac{1}{\sqrt{2 \pi} \sigma_{R_p}} \exp{-\frac{(R_p - \overline{R}_p)^2}{2 {\sigma_{R_p}}^2}}.
\end{equation}
Sampling from this Gaussian results in a set of samples of $R_p$.  The posterior samples of the joint parameter $\sqrt{A_g}R_p$ obtained from MultiNest can be divided by these samples of $R_p$, and squared to produce samples of $A_g$.  By taking the mean and standard deviation of the samples of $A_g$ we can summarize our estimate of the geometric albedo by $A_g = 0.162 \pm 0.007$.  This result is consistent with the maximum albedo of $\max(A_{g}) = 0.148^{+0.027}_{-0.023}$ reported in Esteves et al. \citep{Esteves+etal:2013}.  Repeating this process for the planetary radius of $R_p = 2.2 \pm 0.1 R_J$ estimated by Szab{\'o} et al. \citep{Szabo+etal:2011} resulted in a lower geometric albedo of $A_g = 0.114 \pm 0.013$.

If one is dealing with a non-transiting planet, the same method can be applied to find an estimate of the true radius.
% kkedit  I fixed up the wording here
By assuming a Gaussian probability density function for $A_g$, one can sample values of $A_g$.  During the estimation procedure, MultiNest provides samples of the joint parameter $\sqrt{A_g}R_p$ from the posterior probability.  Given a sample of $A_g$ and a sample of $\sqrt{A_g}R_p$, we can divide the square of the latter by the former and take the square root to obtain a sample of the planetary radius $R_p$.  The resulting distribution can be plotted and examined, or summarized by computing a mean and a variance of the $R_p$ values.  For example, let us assume that for KOI-13b, we have that $\overline{A}_g = 0.15$ with a standard deviation of $\sigma_{A_g} = 0.05$. Our posterior samples of the joint parameter $\sqrt{A_g}R_p$ results in an estimate of the planetary radius of KOI-13b of $R_p = 1.98 \pm 0.31R_J$, which is between the estimates given by Borucki et al. and Szab{\'o} et al., although with a notably higher uncertainty due to the conservative uncertainty in the geometric albedo.

%The model that includes thermal emissions allows one to separately estimate the planetary radius ($R_p$) and the geometric albedo ($A_g$).  For KOI-13b, we found these values to be $R_p = 1.30 \pm 0.10$ and $A_g = 0.50 \pm 0.21$ (Table 5).  Note that the planetary radius is off by nearly $7\sigma$.  In addition, the uncertainty in geometric albedo is almost $50\%$ of the estimated value.  The fact that the eccentricity is estimated to be $e \sim 0.03$ means that it is difficult to distinguish between reflected and thermal emissions.  One would expect that this would translate into large uncertainty values for $R_p$ and $A_g$.  This is certainly true for $A_g$.  However, by constructing a probability landscape for the thermal model applied to KOI-13b in Figure 8 (just as was done for the synthetic data in Section 3) one can see that the planetary radius lies on the edge of a plateau.  For this reason the variance, which is obtained by marginalizing, is perhaps not the best way to quantify uncertainty in this situation.  Clearly, the landscape indicates an upper-bound to the radius which begins falling off around $2R_J$, which is consistent with previous estimates.  By making the radius smaller, one could make the planet hotter (note that the edge of the plateau becomes angled at high temperatures in the $T_d$ vs. $R_p$ subplot) and it is this degeneracy that is responsible for this plateau.

 The eccentric model that includes both reflected light and thermal emissions (Table 6, First column) allows one to separately estimate the planetary radius ($R_p$) and the geometric albedo ($A_g$).  For KOI-13b, we find these values to be $R_p = 0.60 \pm 1.1 R_J$ and $A_g = 0.09 \pm 0.10$.  The algorithm, which in this study is a demonstration ignoring transits, severely underestimated the planetary radius and accounted for this by increasing the dayside temperature drastically to $T_d = 5263 \pm 394$K.  It is expected for the degeneracy to be large in this model since the eccentricity is so low.  A way to potentially combat this problem, other than including transits (which do not occur for non-transiting planets), is to use the following proposed relation between the geometric albedo and the planet-star separation distance \citep{Kane&Gelino:2010}.
\begin{equation}
A_g = \frac{1}{5} \tanh (r-1) + \frac{3}{10}
\end{equation}
where $r$ is the planet-star separation distance in AU, which can be taken as the semi-major axis for orbits with small eccentricities and/or sufficiently short periods where the atmosphere does not have time to significantly change during a single period.  Using this as an approximation for the true albedo gives more reasonable estimates of radii and dayside temperature based on transit observations.  However, precise estimates of planetary radius and dayside temperature remain unattainable with a single bandpass.  One could test between a model that uses this approximation and a model that estimates $A_g$ by computing the evidence.  This approximation also may not be particularly useful for short-period hot-Jupiters since it has a lower limit of $A_g = 0.1477$ and many planets have been found to have lower geometric albedos.

 The model that treated the planetary flux as thermal emission (Table 6, second column) tied for the most probable model, and yielded the best parameter estimates based on the literature values.  This model gave a planetary mass and radius of $M_p = 7.10 \pm 0.38 M_J$ and $R_p = 2.00 \pm 0.61 R_J$, respectively.  This model also gave an orbital inclination and eccentricity of $81.5 \pm 5.7$ degrees and $e = 0.062 \pm 0.005$, which is slightly outside the accepted range for orbital eccentricity given by \citep{Szabo+etal:2011}.  Perhaps most notably, this model gave an estimated dayside temperature of $T_d = 3492.4 \pm 340.4$ K in agreement with the value estimated by \citep{Esteves+etal:2013}.  Since including a reflection component to the planetary flux did not increase the log evidence, we  conclude that thermal emission is the main component of the planetary flux.

As mentioned above, Figure 8 shows the two dimensional slices through the log-likelihood probability taken at the mean parameter values obtained by MultiNest (Table 6, Eccentric Thermal Model).  Just as in the synthetic case, there are many ridges in the probability landscape.  Comparing the two, we see very similar structures in most of the corresponding plots.  This implies that the forward model used both to create the synthetic data and to analyze KOI-13b is reasonable.  Differences between the synthetic case and KOI-13b lie mainly with planetary radius $R_p$ and inclination $\cos i $.  KOI-13b is in a nearly circular orbit, so it is expected that separating thermal and reflected flux with one bandpass will be impossible.  This is evident in the plots involving $R_p$ in the third column where the probability density is much more flat and spread out than in the synthetic case.  Another difference between the synthetic planet and KOI-13b is that the estimated parameter values (seen as black crosses) do not always lie on peaks in the case of KOI-13b.  This is because the estimated parameter values represent mean values which need not reside on a peak since the posterior probability is multimodal.

\begin{figure}[h]	
\centering
\includegraphics[scale=.7]{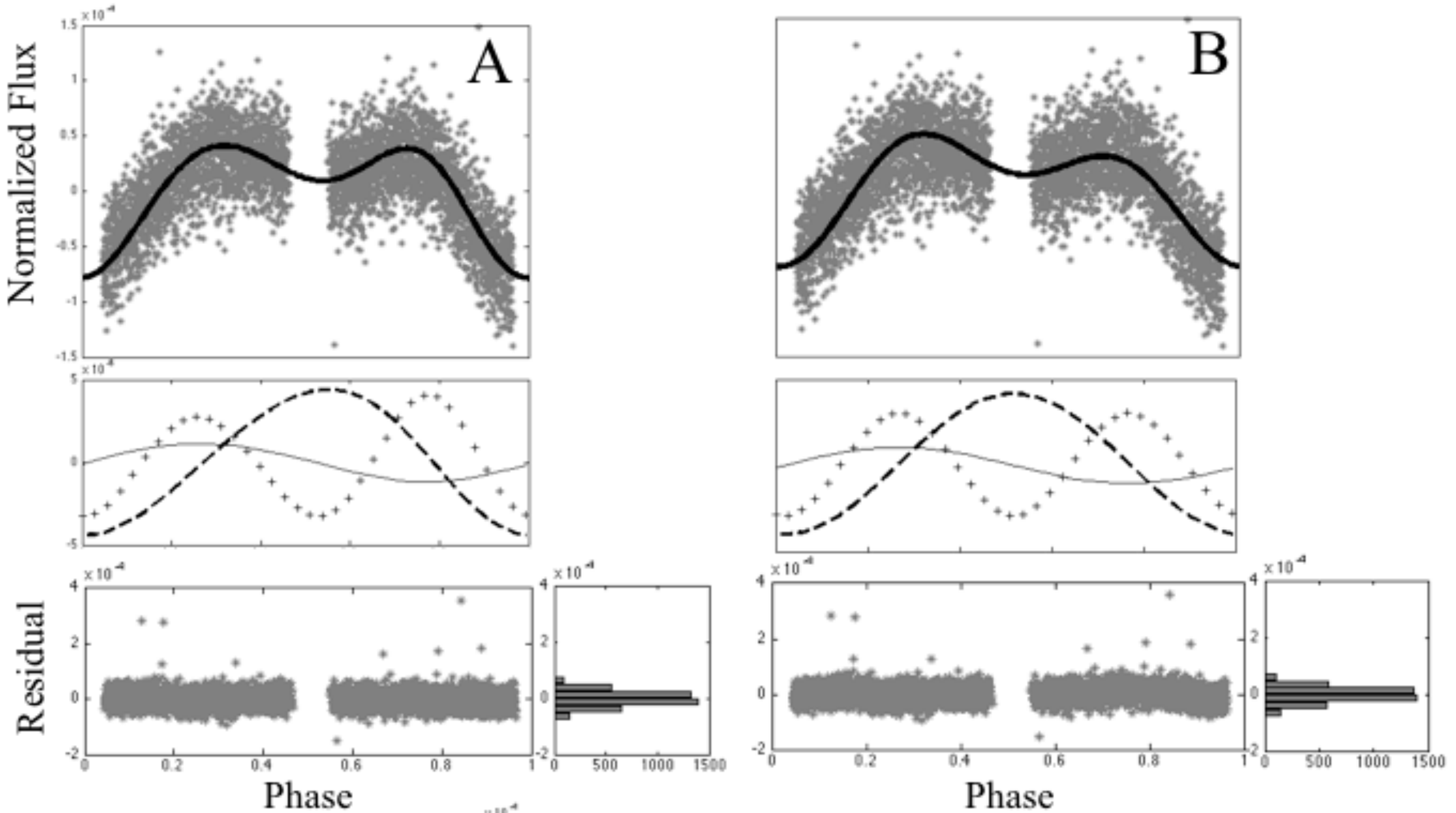}
\caption{Out-of-transit data for KOI-13b including fits for (A) eccentric ($\ln Z = 37\,748 \pm 1.1$; $RSS = 3.45e-06$) and (B) circular ($\ln Z = 37\,703 \pm 1.1$; $RSS = 3.8e-06$) models with all three photometric effects. Each plot shows the contributions from the individual effects below the fit. The dashed line represents the reflected light, + signs represent the ellipsoidal variations, the solid line represents the Doppler beaming.}
\label{Figure:fits}
\end{figure}

\begin{figure}[h]
\centering
\includegraphics[scale = 0.48]{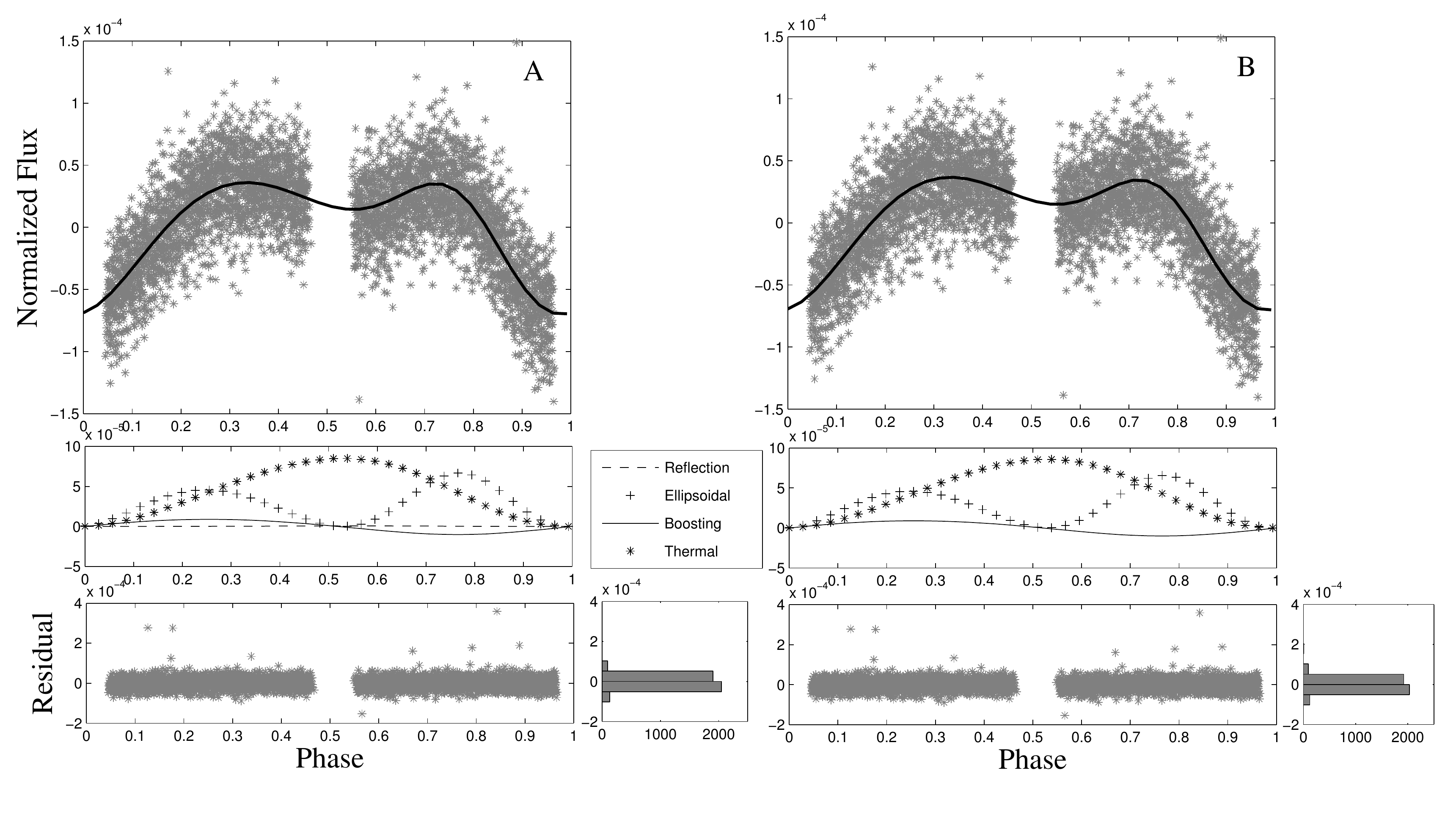}
\caption{Out-of-transit data for KOI-13b including fits for eccentric orbits with (A) reflected light, thermal emission, Doppler beaming, and ellipsoidal variations($\ln Z = 37\,764 \pm 8.3$; $RSS = 3.3686e-06$), and (B) thermal emission, Doppler beaming, and ellipsoidal variations ($\ln Z = 37\,765 \pm 0.9$; $RSS = 3.3674e-06$).}
\label{Figure:fits2}

\end{figure}

\section{Conclusions}{

The EXONEST algorithm was developed to characterize exoplanets based on photometric variations using
Bayesian model selection.  This methodology relies on the Bayesian evidence, which enables one to test a number of models against a dataset.  Algorithms like Nested Sampling and MultiNest enable one to explicitly calculate log evidence values as well as sample from the posterior, which results in parameter estimates as well as the uncertainties in those estimates.  We have demonstrated the abilities of EXONEST to perform model selection using synthetic data as well as exploring the distinguishability of thermal and reflected light.  It was determined using model-generated data that using photometric effects alone, thermal and reflected flux cannot be disentangled for orbits with eccentricity less than 0.3. The problem of disentangling thermal and reflected flux would be resolved by considering two or more different spectral channels.  However, the degree to which this is possible would need to be studied by careful simulations.

 We have also added to results from previous work for the transiting planet KOI-13b and have demonstrated that the photometric effects associated with the out-of-transit data can be used for characterizing many aspects of this planetary system.  It was found that the most favored model to describe the data is a slightly eccentric orbit ($e = 0.062 \pm 0.005$) exhibiting the photometric effects of thermal flux, Doppler boosting, and ellipsoidal variations. The log evidence, $RSS$, and the parameter estimates all suggest that thermal flux represents the main contributor to the planetary flux and that the reflected light plays a minimal role.  We estimate a planetary radius of $2.00 \pm 0.61 R_J$, and a mass of $7.10 \pm 0.38M_J$. Our mass estimates are relatively close to those of Esteves et al. \citep{Esteves+etal:2013}, and Mislis and Hodgkin \citep{Mislis&Hodgkin:2012}, who estimate masses of  $7.93 \pm 0.27M_J$ and $8.30 \pm 1.25 M_J$, respectively.  We also find a dayside temperature of $T_d = 3392.4 \pm 340.4 K$, which is within $1-\sigma$ of that estimated by \citep{Esteves+etal:2013} who estimated the dayside temperature from secondary eclipses as $T_d =3724 \pm 3K$.  It should be noted that since EXONEST estimated the dayside temperature from the phase curve, not the secondary eclipse, it is expected that the uncertainty in $T_d$ should be much larger.  For the models that exclude thermal emission, we used the  lower limit on the planetary radius in conjunction with estimates of the planetary radius obtained from transits to derive estimates of the geometric albedo. Given a planetary radius of $R_p = 1.86 \pm 0.003R_J$ \citep{Borucki+etal:2011}, we estimate a geometric albedo of $A_g = 0.162 \pm 0.007$; whereas a slightly larger planetary radius of $R_p = 2.2 \pm 0.1 R_J$ \citep{Szabo+etal:2011} results in a lower geometric albedo of $A_g = 0.114 \pm 0.013$.  This result seems to be in disagreement with the results from our simulations on model-generated data since we found that the algorithm could not disentangle thermal emission and reflected light up to an eccentricity of $e = 0.3$.  It may be that the simulations are sensitive to the orbital orientation and that some orientations allow for better estimation of the two signals. For eccentric orbits, the argument of periastron $\omega$ dramatically changes the reflected light waveform.  Under the assumption that the temperature of the planet does not change over the course of an orbital period, it could be that certain values of $\omega$ are more suitable for disentangling thermal emission from reflected light since the thermal emission signal would not be significantly altered by changing $\omega$.  Another possible confounding issue is that there may be superrotation occuring in the atmosphere of the planet  \citep{Faigler+etal:2013, Faigler+etal:2014}, which would cause a phase shift in the thermal emission signal. The fits for the eccentric orbit in Figure \ref{Figure:fits}A  indicate a shift in the maximum of the reflected light to the right of phase $0.5$. The fits for thermal emission in Figure \ref{Figure:fits2} also show the same shift in the maximum of the thermal flux.  While this shift is attributed to the slight eccentricity detected by EXONEST, it could also be a sign of superrotation, which could be modeled and tested against the existing suite of models in Table $4$.

In the future, we will be extending the EXONEST algorithm by including models of transits and secondary eclipses\footnote{Since the initial submission of this manuscript, we have succeeded in extending the EXONEST algorithm to accomodate transits and secondary eclipses, and are currently completing a study of these results.}. This should enable reflected light and thermal emissions to be better disentangled and thus more precisely estimated.  It should also allow one to better estimate the parameters relevant to the other photometric effects as transits and secondary eclipses hold information on the planetary radius, dayside temperature, and orbital inclination.  Also, instead of working only with uniform distributions, we can assign Jeffrey's priors \citep{Sivia&Skilling:2006} to scale parameters, or take into account newly-obtained information about exoplanets that either limit parameter ranges or better quantify expected parameter values \citep{Kipping:2013,Kipping:2014,Kane&Gelino:2012,Gregory&Fischer:2010}.  We also plan to make EXONEST publicly available in both MATLAB and Python allowing for the incorporation of plug-and-play models.  EXONEST not only holds promise for detecting non-transiting exoplanets from photometric effects alone, but it may also serve as a verification technique since each of these effects can be treated and tested for both separately and jointly.
}

\begin{figure}
\includegraphics[scale=.58]{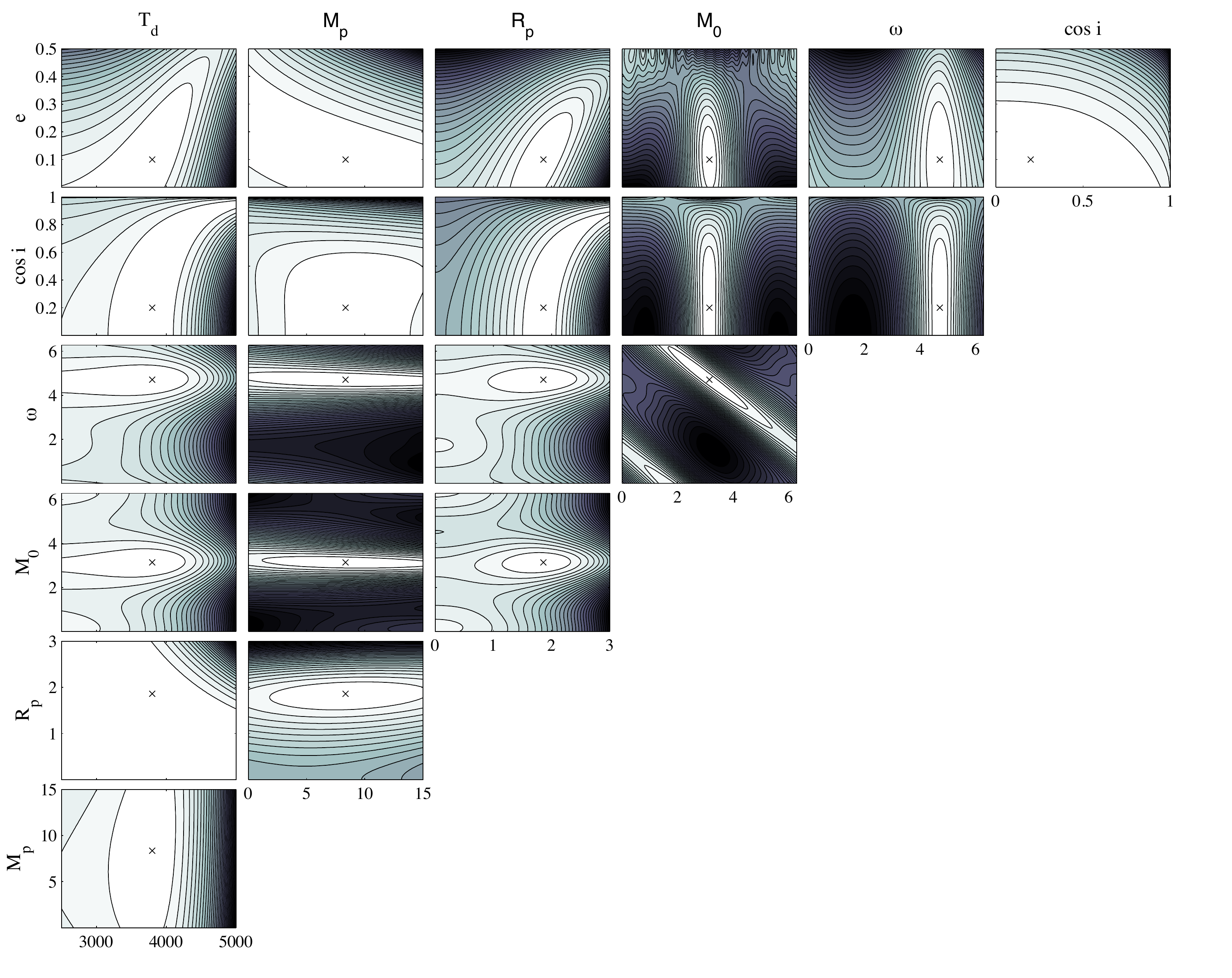}
\caption{Synthetic Log-Likelihood landscapes for pairs of model parameters.  Slices were taken through parameter space at the true parameter values.  The correct solution for this case is given by $\left\{ e = 0.1, \cos i = 0.2, \omega = \frac{3\pi}{2}, M_0 = \pi, R_p = 1.86, M_p = 8.35, T_d = 3800 , A_g = 0.15 \right\}$.  Plots involving $\cos i$, $T_d$, and $M_p$ are seen to be relatively flat implying more uncertainty in those parameters.  The argument of periastron ($\omega$) and initial mean anomaly ($M_0$) show very sharp peaks and are thus more accurately estimated.}
\end{figure}

\begin{figure}[h]
\includegraphics[scale=.58]{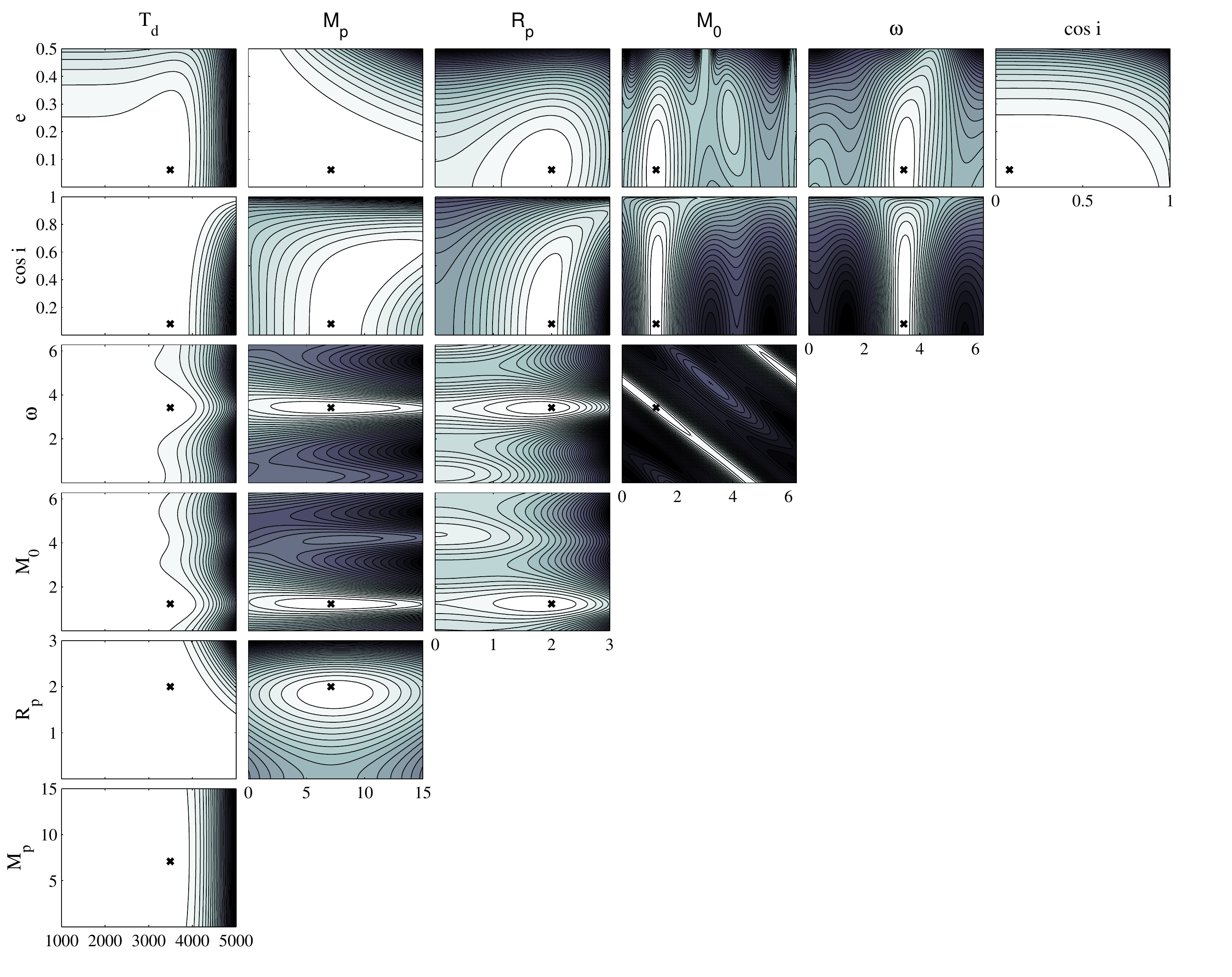}
\caption{Log-Likelihood landscapes for pairs of model parameters including  applied to KOI-13b. Slices were taken through parameter space at the mean parameter values (labeled by black crosses) obtained from MultiNest - $\left\{ e = 0.062, \cos i = 0.08, \omega = 3.42, M_0 = 1.23, R_p = 2.0, M_p = 7.1, T_d = 3492 \right\}$.  Most of the slices through the probability landscape of KOI-13b look similar to that of the synthetic planet implying that the forward model is realistic. }
\end{figure}

%%% The following command ends your manuscript. LaTeX will ignore any text
%%% that appears after it.
\bibliographystyle{plainnat}
\bibliography{mybib}

\end{document}